\documentclass[sigconf]{acmart}

\makeatletter
\def\@ACM@checkaffil{
    \if@ACM@instpresent\else
    \ClassWarningNoLine{\@classname}{No institution present for an affiliation}%
    \fi
    \if@ACM@citypresent\else
    \ClassWarningNoLine{\@classname}{No city present for an affiliation}%
    \fi
    \if@ACM@countrypresent\else
        \ClassWarningNoLine{\@classname}{No country present for an affiliation}%
    \fi
}
\makeatother

\AtBeginDocument{%
  }

\copyrightyear{2025}
\acmYear{2025}
\setcopyright{acmlicensed}\acmConference[ISCA '25]{Proceedings of the 52nd Annual International Symposium on Computer Architecture}{June 21--25, 2025}{Tokyo, Japan}
\acmBooktitle{Proceedings of the 52nd Annual International Symposium on Computer Architecture (ISCA '25), June 21--25, 2025, Tokyo, Japan}
\acmDOI{10.1145/3695053.3731070}
\acmISBN{979-8-4007-1261-6/2025/06}

\usepackage{multirow}
\usepackage{subfigure}
\usepackage{bbding}
\usepackage{mathtools}

\usepackage{algorithm}
\usepackage{algpseudocode}
\algtext*{EndFor}
\algtext*{EndFunction}
\algtext*{Until}
\algtext*{EndIf}
\algtext*{EndWhile}
\usepackage{amsmath}
\usepackage{array}
\usepackage{tablefootnote}

\usepackage{geometry}

\usepackage{soul}
\usepackage{siunitx}
\usepackage{multicol}

\newcommand{\yaot}[1]{\textcolor{black}{#1}} 
\newcommand{\yao}[1]{\textcolor{black}{#1}}

\newcommand{\mm}[1]{\textcolor{black}{#1}}
\newcommand{\mmi}[1]{\textcolor{black}{#1}}

\begin{document}

\begin{abstract}

Temporal prefetching shows promise for handling irregular memory access patterns, which are common in data-dependent and pointer-based data structures. Recent studies introduced on-chip metadata storage to reduce the memory traffic caused by accessing metadata from off-chip DRAM. However, existing prefetching schemes struggle to efficiently utilize the limited on-chip storage. \mmi{An alternative solution, software indirect access prefetching, remains ineffective for optimizing temporal prefetching.}

In this work, we propose Prophet---a hardware-software \yaot{co-designed} framework that leverages profile-guided methods to optimize metadata storage management. Prophet profiles programs using counters instead of traces, injects hints into programs to guide metadata storage management, and dynamically tunes these hints to enable the optimized binary to adapt to different program inputs. Prophet is designed to coexist with existing hardware temporal prefetchers, delivering efficient, high-performance solutions for frequently \yaot{executed} workloads while preserving the original runtime scheme for less frequently \yaot{executed} workloads. Prophet outperforms the state-of-the-art temporal prefetcher, Triangel, by \mmi{14.23\%}, effectively addressing complex temporal patterns where prior profile-guided solutions fall short (only achieving 0.1\% performance gain). Prophet delivers \yaot{superior} performance across all evaluated workload inputs, introducing negligible profiling, analysis, and instruction overhead.

\end{abstract}



\title{Profile-Guided Temporal Prefetching}

\author{Mengming Li}
\affiliation{%
  \institution{HKUST}}
\email{mengming.li@connect.ust.hk}

\author{Qijun Zhang}
\affiliation{%
  \institution{HKUST}}
\email{qzhangcs@connect.ust.hk}

\author{Yichuan Gao}
\affiliation{%
  \institution{Intel}
}
\email{yichuan.gao@intel.com}

\author{Wenji Fang}
\affiliation{%
  \institution{HKUST}
}
\email{wfang838@connect.ust.hk}

\author{Yao Lu}
\affiliation{%
  \institution{HKUST}
}
\email{yludf@connect.ust.hk}

\author{Yongqing Ren}
\affiliation{%
  \institution{Intel}
}
\email{yongqing.ren@intel.com}

\author{Zhiyao Xie}
\affiliation{%
  \institution{HKUST}
}
\email{eezhiyao@ust.hk}
\authornote{Corresponding Author}

\maketitle
\pagestyle{plain}

\section{Introduction}
\label{sec:intro}

Data prefetching, a widely studied technique for addressing the ``memory wall'' \cite{wulf1995hitting}, has been extensively researched to enhance processor performance. Among various data prefetching techniques \cite{bakhshalipour2019evaluation, mittal2016survey, jouppi1990improving, he2022dsdp, hur2006memory, baer1991effective, dahlgren1995effectiveness, kim1997stride, somogyi2006spatial, kim2016path, michaud2016best, navarro2022berti, bakhshalipour2019bingo, bera2019dspatch, pugsley2014sandbox, shevgoor2015efficiently, ishii2009access, jiang2022merging, jain2013linearizing, nesbit2004data, wu2019efficient, wenisch2009practical, wu2019temporal, bakhshalipour2018domino, panda2023clip,ainsworth2024triangel,wu2021practical,alecto}, temporal prefetching \cite{jain2013linearizing, nesbit2004data, wu2019efficient, wenisch2009practical, wu2019temporal, bakhshalipour2018domino, ainsworth2024triangel} shows particular promise for addressing irregular memory access patterns, which often arise from indirect/data-dependent memory accesses and pointer-based data structures. Temporal prefetchers typically require significant metadata storage to record correlations between memory addresses, making efficient metadata storage design vitally important. 
Recent studies \cite{wu2019temporal, wu2021practical, ainsworth2024triangel} propose relocating metadata storage from off-chip DRAM \cite{jain2013linearizing, nesbit2004data, wu2019efficient, wenisch2009practical, bakhshalipour2018domino} to the on-chip metadata table within LLC to reduce memory traffic. Since on-chip storage is a limited resource, \yaot{efficient management of} the metadata table becomes even more critical.

\textbf{Hardware temporal prefetcher.} Existing hardware temporal prefetchers \cite{wu2019temporal, ainsworth2024triangel, wu2021practical} employ various techniques \yaot{for metadata table management}, such as training data filtering, replacement policies, and resizing. However, they fail to balance performance gains with storage overhead. For example, Triage \cite{wu2019temporal} adopts an advanced replacement policy (e.g., Hawkeye \cite{jain2016back}) that incurs a 13~KB storage overhead but results in only a 0.25\% performance gain. The state-of-the-art temporal prefetcher, Triangel \cite{ainsworth2024triangel}, introduces additional techniques like training data filtering to improve metadata table management. \yaot{However}, according to its \yaot{own} ablation study \cite{ainsworth2024triangel}, Triangel’s performance gain \yao{mostly} comes from aggressive prefetching \yaot{instead of its} metadata table management, 
\yaot{which actually incurs 90\% of the} storage overhead. 

\begin{figure}[t]
\centering
\includegraphics[width=.99\linewidth]{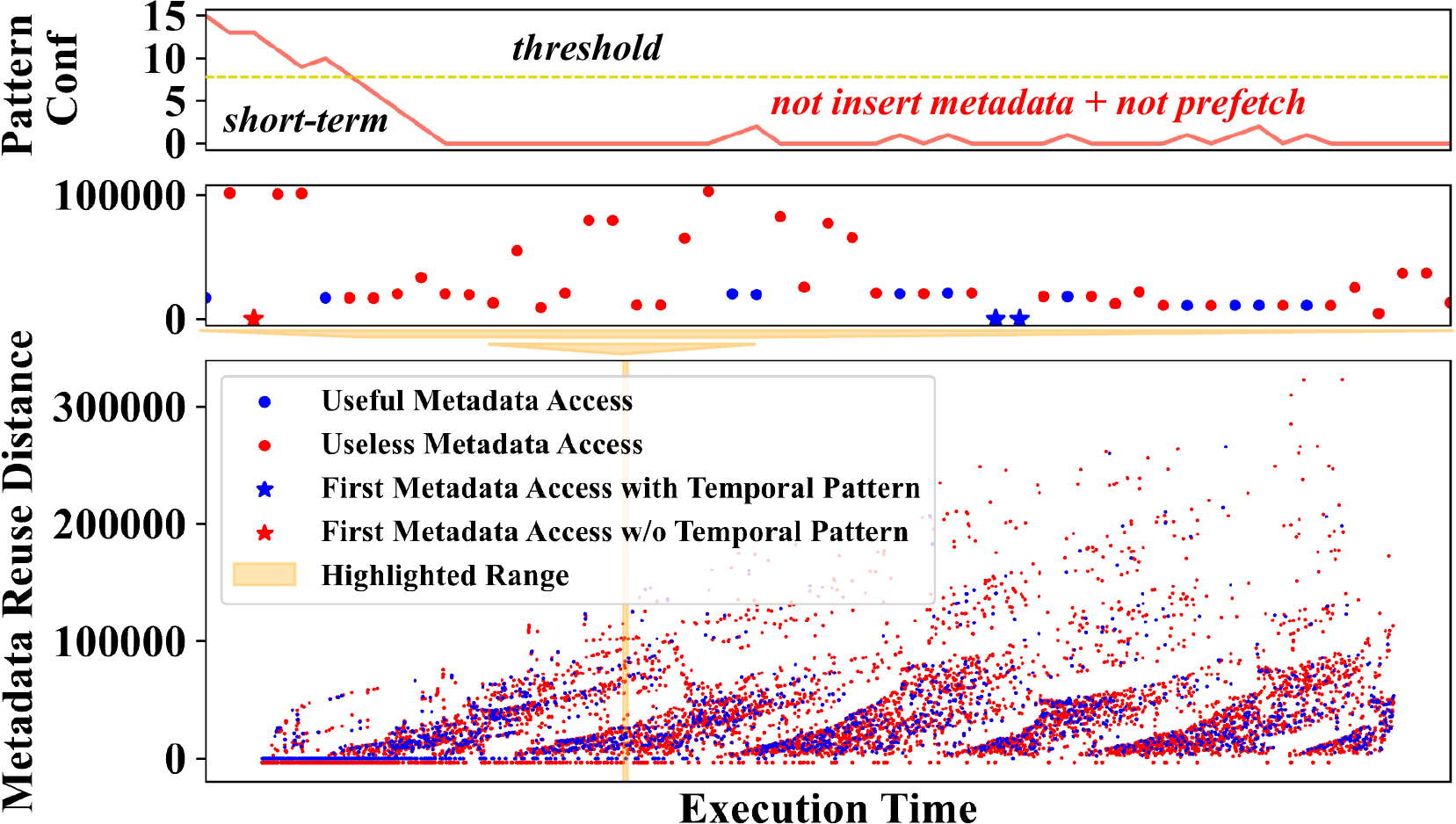}
\vspace{-.1in}
\caption{\mmi{The bottom figure shows a metadata access pattern: 1) Blue/Red \emph{dots} are metadata accesses that result in useful/useless prefetches; 2) Blue/Red \emph{stars} represent first metadata access with/without temporal patterns. Their corresponding metadata should/should not be inserted in the metadata table. The top figure shows how Triangel \cite{ainsworth2024triangel} applies its \emph{PatternConf} to the highlighted metadata access pattern.}}
\vspace{-.15in}
\label{fig:motiv}
\end{figure}


Ideal metadata table management involves two key requirements: (1) storing all metadata that \yaot{contributes} to useful prefetches in the metadata table; and (2) filtering out metadata that does not contribute to useful prefetches. However, existing temporal prefetchers \cite{wu2019temporal, wu2021practical, ainsworth2024triangel} fail to balance these two requirements. They either store many invalid metadata entries (e.g., \mmi{no insertion policy for} Triage \cite{wu2019temporal}) or incorrectly filter \yaot{out} metadata entries that could result in useful prefetches (e.g., \mmi{overly conservative insertion policy for} Triangel \cite{ainsworth2024triangel}).


\mmi{To investigate why Triangel's metadata management is inefficient, Figure~\ref{fig:motiv} shows a metadata access pattern\footnote{\mmi{This pattern is derived from a hardware temporal prefetcher with an unlimited metadata table size and no insertion policy.}}$^,$\footnote{\mmi{This pattern is extracted from a frequently accessed instruction in omnetpp, a workload where Triangel shows limited effectiveness.}} and analyzes how Triangel applies its \textit{PatterConf} to the pattern. The 4-bit \textit{PatternConf} evaluates whether future memory accesses exhibit temporal patterns. Triangel utilizes past short-term data to update the \textit{PatternConf}. Specifically, useful metadata accesses (blue dots) increase the \textit{PatternConf}, while useless metadata accesses (red dots) decrease it. When \textit{PatternConf} falls below a threshold, indicating the absence of temporal patterns in the future, Triangel will completely disable metadata insertion (no insertion for stars in Figure~\ref{fig:motiv}).}


\mmi{We have two observations for Figure~\ref{fig:motiv}: (1) The metadata access patterns of temporal prefetching are highly variable, characterized by interleaved useful (blue) and useless (red) metadata accesses and large metadata reuse distance variance. (2) Triangel does not adapt to such dynamic variance of temporal patterns. As shown in Figure~\ref{fig:motiv}, \textit{PatterConf} drops to 0 due to the occurrence of many red dots in a short term. As a result, Triangel incorrectly rejects the insertion of subsequent interleaved blue stars (first metadata access with temporal pattern). Similar challenges also affect other metadata management strategies, as we will cover in Section~\ref{subsec:htp}.}

Purely hardware-based temporal prefetching fails to handle the dynamic variance of metadata accesses due to two main reasons: (1) they lack visibility into future program behavior, and (2) while long-term execution data could help mitigate this problem, storing and analyzing such data would introduce significant performance and storage overhead.

\textbf{Profile-guided temporal prefetcher.} To address the above challenges, profile-guided techniques are promising by leveraging comprehensive program execution data to guide metadata table management strategies. 
However, existing profile-guided solutions \cite{zhang2024rpg2, jamilan2022apt, ainsworth2017software, khan2021dmon} for prefetching are \emph{ineffective for most temporal patterns.} These solutions primarily focus on inserting software prefetch instructions for indirect memory accesses \emph{where the prefetch kernel follows a regular stride pattern}. \yao{They struggle to handle more complex irregular patterns, such as pointer-chasing accesses and indirect memory accesses with complicated prefetch kernels.} 
\mmi{This limitation arises primarily because many irregular patterns involve long-chain dependencies \cite{ayers2020classifying}, and computing dependent addresses along the chain significantly impacts prefetching timeliness.}

\begin{figure}[t]
\centering
\includegraphics[width=.99\linewidth]{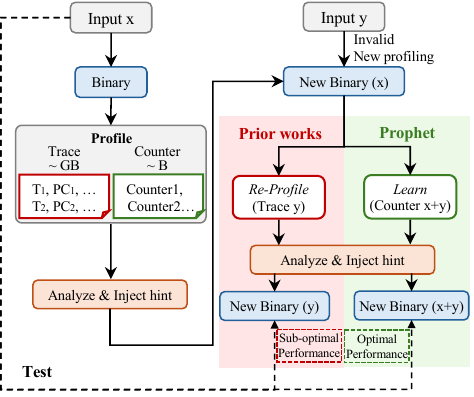}
\vspace{-.1in}
\caption{Comparison between Prophet and prior profile-guided solutions. Prophet is \emph{lightweight} as it only uses counters for profiling. Prophet can integrate counters from multiple inputs, enabling it to \emph{adapt to} varying program inputs.}
\vspace{-.15in}
\label{fig:insight}
\end{figure}

\textbf{Our solution named Prophet.} \mmi{Driven by our analysis, we propose \texttt{Prophet}\footnote{Prophet is open-sourced at: \url{https://github.com/hkust-zhiyao/Prophet}.}---a hardware-software co-designed framework, which maintains the metadata table within hardware temporal prefetchers while leveraging profile-guided methods to improve metadata table management. Prophet addresses aforementioned challenges without requiring significant hardware overheads: it offloads hardware-intensive tasks (e.g., data analysis) to software and injects hints into programs to guide the metadata table \emph{insertion policy}, \emph{replacement policy}, and \emph{resizing operations}.}


\mm{Beyond the scope of temporal prefetching, Prophet introduces an innovative, efficient profile-guided solution that overcomes key challenges in traditional profile-guided methods, as outlined below.}
\begin{itemize}
    \item \textbf{Adaptable.} \mm{We investigate why profile-guided optimizations often struggle with varying program inputs and why hints derived from one input may not apply to others. Based on these insights, we enable Prophet to integrate profiling data from multiple program inputs. Leveraging the aggregated data, Prophet can generate a single optimized binary that adapts effectively across these inputs, as shown in Figure~\ref{fig:insight}.}
    \item \textbf{Lightweight.} Prophet leverages \mmi{Performance Monitoring Unit (PMU)} counters instead of traces to profile programs, offering two major benefits: (1) it avoids the significant performance and storage overhead linked to trace-based profiling; (2) it is readily applicable to current architectures without requiring additional memory trace systems.
    \item \textbf{Compatible.} Prophet can co-exist with existing hardware temporal prefetchers. Prophet offers high-performance yet efficient solutions for frequently \yao{executed} workloads while maintaining the original runtime solution (e.g., Triangel \cite{ainsworth2024triangel}) for rarely \yao{executed} workloads.
\end{itemize}

We evaluate Prophet on applications representative of temporal patterns, which are commonly used in prior studies \cite{wu2019efficient,wu2019temporal,wu2021practical,ainsworth2024triangel}. Across all applications, Prophet outperforms the state-of-the-art software indirect memory access prefetching schemes, RPG$^2$ \cite{zhang2024rpg2} by \mmi{34.48\%} and hardware temporal prefetcher, Triangel \cite{ainsworth2024triangel} by \mmi{14.23\%}. 
This performance gain is driven by Prophet’s efficient metadata table management, which not only significantly enhances prefetching coverage but also maintains high prefetching accuracy. Extensive evaluations demonstrate that Prophet can adapt to varying program inputs while introducing negligible profiling, analysis, and instruction overheads.

\section{Background}


\subsection{Hardware Temporal Prefetching}
\label{subsec:htp}



\begin{figure}[t]
\centering
\includegraphics[width=.8\linewidth]{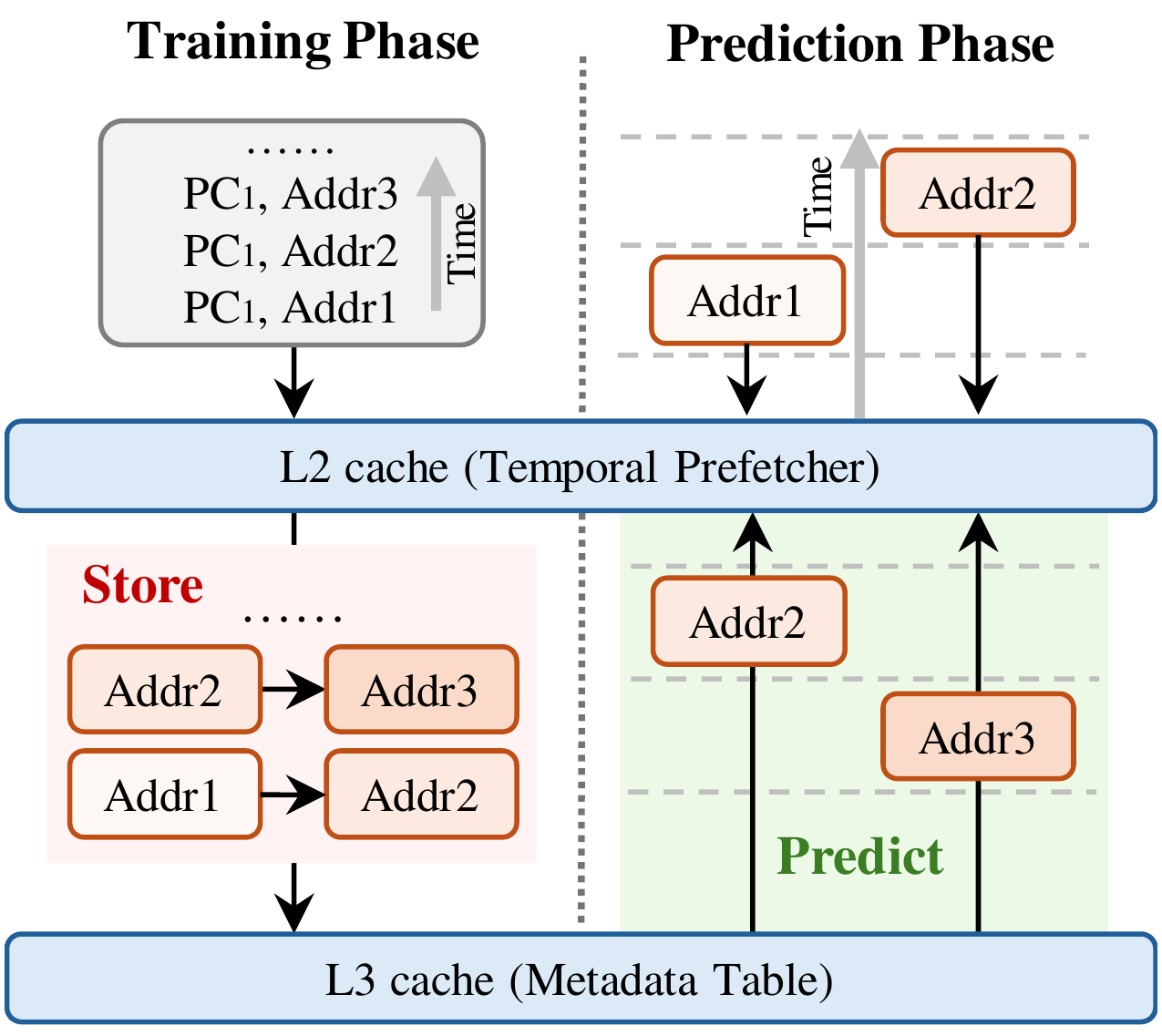}
\vspace{-.1in}
\caption{\mmi{A general framework of temporal prefetching.}}
\vspace{-.2in}
\label{fig:hw_tp}
\end{figure}

\mmi{As shown in Figure~\ref{fig:hw_tp}, the core idea of hardware temporal prefetching \cite{jain2013linearizing, nesbit2004data, wu2019efficient, wenisch2009practical, wu2019temporal, bakhshalipour2018domino, ainsworth2024triangel, wu2021practical} is to record previously accessed memory addresses and their correlations, referred to as metadata. When recorded addresses are re-accessed, the prefetcher can use the correlation information to predict future memory accesses.} To record metadata of various temporal patterns, hardware temporal prefetching algorithms require significant data storage. Early temporal prefetchers \cite{jain2013linearizing, nesbit2004data, wenisch2009practical, bakhshalipour2018domino, wu2019efficient} utilized DRAM to store the metadata. However, fetching metadata from DRAM consumes a substantial amount of memory bandwidth that could otherwise be used for demand memory accesses. To address this issue, Triage \cite{wu2021practical, wu2019temporal} recently proposed storing metadata in a Markov table that shares space with LLC, eliminating the need to load metadata from off-chip memory. However, on-chip storage is precious, so temporal prefetchers must operate with a limited metadata table size.

To efficiently utilize on-chip memory, Triage \cite{wu2021practical, wu2019temporal} and the state-of-the-art temporal prefetcher, Triangel \cite{ainsworth2024triangel}, introduce several techniques for managing the metadata table:

\subsubsection{Insertion policy (training data filtering).} Not every demand request should be used to train the temporal prefetcher and trigger metadata insertion. To improve the utilization of the metadata table, temporal prefetchers should focus on handling demand requests that exhibit \emph{solvable} temporal patterns. The ``solvable'' indicates these patterns repeat in a short enough sequence to fit in the metadata table. Ideally, the temporal prefetcher should reject metadata insertion for demand requests that do not meet this criterion.
    
The first on-chip temporal prefetching scheme, Triage, \yao{does not} implement any insertion \yao{(i.e. filtering)} policy. The state-of-the-art temporal prefetcher, Triangel, leverages \textit{PatternConf} and \textit{ReuseConf} to identify and \yao{filter out} demand requests that do not fall within the capabilities of temporal prefetchers. The \textit{PatternConf} \yao{checks} whether demand requests from a memory instruction exhibit a temporal pattern. The \textit{ReuseConf} further evaluates whether the temporal pattern can fit the metadata table.

\textbf{Inefficiency of insertion policy.} According to Triangel’s ablation study \cite{ainsworth2024triangel}, filtering out demand requests \yao{with \textit{PatternConf} and \textit{ReuseConf} only yields} marginal overall performance improvements and even degrades the performance for many applications. This limitation stems from the inaccuracy of these metrics \yao{(i.e., \textit{PatternConf} and \textit{ReuseConf})}, as illustrated in Figure~\ref{fig:motiv}. \yao{As a result,} \mm{inaccurate filtering discards metadata that could lead to useful prefetches.}

\mm{\textbf{Insertion policy in Prophet.} Prophet implements a more accurate insertion policy, filtering out only metadata that is highly unlikely to originate from temporal patterns. This is accomplished by identifying PCs whose overall prefetching accuracy falls below an extremely low threshold, a metric easily obtained during the profiling stage. Unlike hardware temporal prefetchers, which rely on short-term data to guide future program execution, the metrics gathered during profiling (e.g., prefetching accuracy) reflect the actual behavior of programs, resulting in more precise decisions.}

\subsubsection{Replacement Policy.} When the metadata table capacity is insufficient to accommodate a new entry, the temporal prefetcher must decide which existing metadata entry to evict. Triage \cite{wu2019temporal} observed that only a small fraction of metadata is frequently reused. As a result, it employs an advanced replacement policy (e.g., Hawkeye \cite{jain2016back}) to evict metadata entries that are less likely to be accessed in the future, thereby enhancing the utilization of metadata tables. However, according to Triangel \cite{ainsworth2024triangel}, this replacement policy provides only marginal performance gains, achieving a speedup of less than 0.25\%. As a result, Triangel replaces Hawkeye in Triage with a simpler replacement policy---SRRIP \cite{jaleel2010high}---to balance storage overhead with performance gains. 

\mm{\textbf{Inefficiency of replacement policy.} (1) The reuse distance of metadata entries varies significantly, as shown in Figure~\ref{fig:motiv}. These replacement policies solely focus on predicting reuse distances, making them inefficient for handling such high variance \cite{song2022thermometer}. (2) They focus solely on increasing \yao{hits in metadata table} without considering if metadata hits further result in useful prefetches.}

\mm{\textbf{Replacement policy in Prophet.} Prophet enhances the replacement policy by incorporating prefetching accuracy as an additional metric for selecting victim entries in the metadata table. After filtering out metadata through the insertion policy, Prophet assigns different priority levels to the remaining metadata based on the prefetching accuracy. Metadata entries that are less likely to generate useful prefetch requests are assigned lower priority levels, making them prioritized candidates for replacement. Like the insertion policy, Prophet obtains prefetching accuracy during the profiling stage, ensuring precise and effective management.}






\subsubsection{Resizing.}
\label{sec:resizing}
The metadata table shares space with the LLC, requiring a careful balance in space allocation to avoid compromising the LLC’s ability to serve regular demand requests. Previous schemes address this trade-off using methods like Bloom Filters \cite{bloom1970space} or Set Dueller \cite{ainsworth2024triangel}. Triage employs a Bloom Filter to calculate the effective entries in the metadata table, but tracking approximately 200,000 entries incurs a storage overhead exceeding 200~KB. \mmi{To mitigate this, Triangel introduces the Set Dueller, which uses a small subset of cache sets to model the full-size regular LLC and Markov table. The Set Dueller works by simulating various partitioning configurations for the cache and the Markov table, evaluating their respective hit rates. After a defined window, the Set Dueller selects the configuration that maximizes the hit rate. This approach reduces resizing overhead to approximately 2~KB by tracking only selected cache sets instead of the entire metadata table.}

\textbf{Inefficiency of resizing.} Similar to insertion and replacement policies, \mm{the hit rate between LLC and metadata table} sampled by Triangel's resizing sometimes fails to accurately predict future metadata table requirements. For example, in workloads like omnetpp and mcf, we observe that Triangel's resizing often chooses overly conservative metadata table sizes. Although this improves prefetching accuracy, it reduces prefetching coverage, ultimately damaging overall system performance. As a result, Set Dueller provides only limited performance benefits for Triangel.

\mm{\textbf{Resizing in Prophet.} \mmi{Prophet leverages profile-guided techniques to implement a Bloom-Filters-like method, which helps maintain precision while avoiding significant storage overhead introduced by runtime approaches. Specifically, Prophet allocates storage to the metadata table based on the peak metadata usage observed during the profiling stage. This approach is adopted because resizing provides only marginal performance gains (Section~\ref{subsec:breakdown}), while incorrect resizing can significantly degrade performance.}}

\subsection{Software Indirect Access Prefetching}
\label{subsec:drawback_soft}

\begin{figure*}[t]
\vspace{-.25in}
\centering
\includegraphics[width=.75\linewidth]{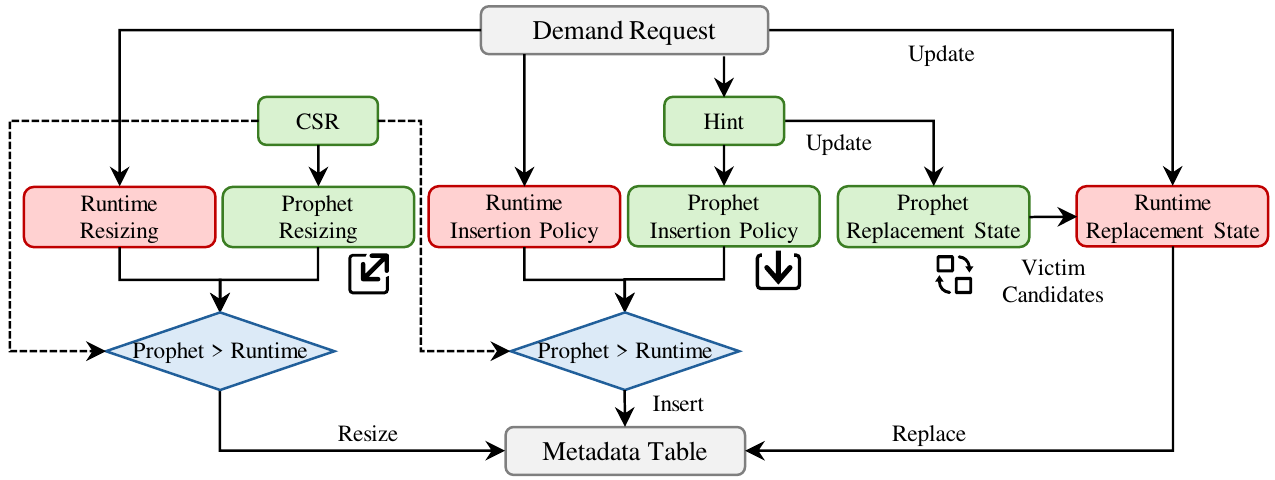}
\vspace{-.1in}
\caption{Prophet architecture overview. Prophet coexists with hardware temporal prefetchers by sharing the same metadata table but leveraging more accurate profile-guided methods for metadata table management.}
\vspace{-.15in}
\label{fig:structure_overview}
\end{figure*}

Software indirect access prefetching \cite{zhang2024rpg2, jamilan2022apt, ainsworth2017software, khan2021dmon, callahan1991software, chen1991data, gornish1990compiler} predicts indirect memory accesses by inserting prefetch instructions into programs. Indirect memory accesses include prefetch kernel accesses (e.g., \textit{b[i])} and data-dependent memory accesses linked to that prefetch kernel (e.g., \textit{a[b[i]]}).

Software indirect access prefetching primarily targets situations where the prefetch kernel follows a stride pattern, such as when the accesses for a[b[i]] occur in loops with indices \textit{i, i + d, i + 2d}, and so on. These schemes generally follow a three-step process: (1) identifying the prefetch kernel; (2) calculating the prefetch distance between the insertion position of software prefetch instructions and the prefetch kernel; and (3) inserting the software prefetch instructions into original programs. The most critical aspect is calculating the prefetching distance, as the position of prefetch instructions determines the prefetching timeliness. 

Based on the methods used to calculate the prefetching distance, previous works can be categorized into static solutions \cite{callahan1991software, chen1991data, gornish1990compiler} and profile-guided optimization (PGO) solutions \cite{zhang2024rpg2, jamilan2022apt, ainsworth2017software, khan2021dmon}. In static solutions, programmers manually determine where to insert software prefetch instructions. However, due to the complexity of indirect memory accesses, manual insertion is prone to errors and may cause performance slowdowns. Profile-guided solutions overcome the limitations of static approaches. They profile running programs to automatically identify optimal positions for inserting prefetch instructions. Experimental results \cite{zhang2024rpg2,jamilan2022apt} indicate that these approaches can significantly improve system performance on certain graph benchmarks, such as CRONO \cite{ahmad2015crono}.


\textbf{Inefficiency of software indirect access prefetching.} Existing software indirect access prefetching schemes are effective only for a narrow subset of indirect memory accesses where the prefetch kernel follows a regular stride pattern. They fail to address most irregular patterns \mmi{effectively}, including complex indirect accesses (e.g., prefetch kernel without stride patterns) and pointer-chasing accesses. \mmi{This limitation arises primarily because many irregular patterns involve long-chain dependencies \cite{ayers2020classifying}, and computing dependent addresses along the chain significantly affects the prefetching timeliness.}

To validate their limitations, in Section~\ref{subsec:perf}, we evaluate the state-of-the-art PGO-based scheme, RPG$^2$ \cite{zhang2024rpg2}, on representative SPEC CPU workloads commonly used in temporal prefetching studies \cite{wu2019temporal, wu2021practical, ainsworth2024triangel}. Normalized to a baseline without a temporal prefetcher, RPG$^2$ achieves only a 0.1\% performance improvement, significantly lower than its gains on graph benchmarks. We observe that this underperformance is due to the complexity of indirect memory accesses in the evaluated workloads. For instance, in mcf, the index of a prefetch kernel is derived through a series of logical operations and multi-step arithmetic computations.

\mm{\textbf{Solution in Prophet.} Prophet is applicable to all types of temporal patterns because it gets rid of software prefetch instruction. Prophet only guides the execution of hardware structures with hints. By preserving the core functions of the metadata table, Prophet maintains the ability to handle all temporal patterns like hardware temporal prefetchers.}

\section{Overview}
\label{sec:design}

\subsection{Architecture Overview}
\label{subsec:structure}


\yao{This section provides an overview} of Prophet architecture, which is \textbf{compatible} with existing hardware temporal prefetchers. \mmi{For the metadata format, Prophet packs 12 compressed metadata entries inside each 64-byte cache line, with each metadata entry containing a 10-bit tag and a 31-bit target address. For the metadata table management,} Prophet offers efficient, high-performance solutions for frequently \yao{executed} workloads while maintaining the original runtime solution (e.g., Triangel) for rarely \yao{executed} workloads. As shown in Figure~\ref{fig:structure_overview}, Prophet consists of three components: profile-guided insertion policy, profile-guided replacement policy (along with its associated replacement states), and profile-guided resizing operations. These profile-guided components rely on two types of information granularity: \yao{application-level and PC-level}. The \yao{application-level} information is embedded in the Control and Status Register (CSR). The \yao{PC-level} information is accompanied by demand requests, referred to as \textit{Hint} in \yao{Figure~\ref{fig:structure_overview}}. Next, we will overview these components:


\textbf{Prophet Insertion Policy:} This policy determines whether to train the temporal prefetcher with demand requests and insert their associated metadata into the table by checking the hint information carried by the demand requests. We disable the Runtime Insertion Policy when the Prophet Insertion Policy is enabled. Prophet’s building blocks are activated through CSR manipulation. Specifically, after profiling the program, a CSR manipulation instruction is inserted at the beginning of the binary to enable Prophet.

\textbf{Prophet Replacement Policy:} This policy assigns replacement priorities to demand requests, with the priority information carried by hints within demand requests. Upon inserting new metadata, we record the priority information from hints into the Prophet Replacement State. During the replacement process, the Prophet Replacement Policy first generates candidate victims for the Runtime Replacement Policy, which then chooses the final victim.

\textbf{Prophet Resizing:} \yao{This policy} retrieves the target metadata table size from the CSR and allocates LLC space to the metadata table at the beginning of program execution. Similar to the Prophet Insertion Policy, we disable the Runtime Resizing when Prophet Resizing is enabled.

\mm{\textbf{Compatibility.} In our framework, Prophet coexists with hardware temporal prefetchers by sharing the same metadata table but leveraging more accurate profile-guided methods for metadata table management. Additionally, Prophet reuses the runtime solution's replacement states, integrating both reuse distance and prefetching accuracy into its replacement policy for enhanced effectiveness. Programmers can switch between Prophet and the hardware temporal prefetcher based on application execution frequency and the trade-off between Prophet's performance gains and its impact on DRAM traffic (Section~\ref{subsec:breakdown}).}

\subsection{Process Overview}
\label{subsec:process}

\begin{figure}[t]
\centering
\includegraphics[width=.99\linewidth]{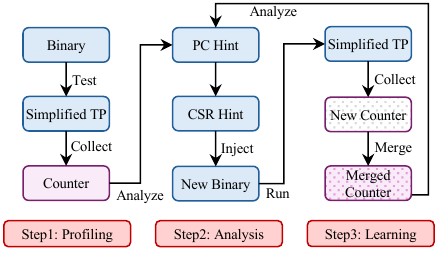}
\vspace{-.1in}
\caption{Prophet process overview. Step1: Prophet leverages the PMU to gather counters related to the temporal prefetcher’s performance. Step2: Prophet analyzes the collected counters to generate hints and then injects hints into the original binaries. Step 3: Prophet samples and learns counters across different program inputs.}
\vspace{-.2in}
\label{fig:process_overview}
\end{figure}

Figure~\ref{fig:process_overview} outlines the process flow of Prophet, comprising three steps: Profiling, Analysis, and Learning. Next, we overview the operations involved in each step of Prophet's process:

\textbf{Step1: Profiling \mmi{(Section~\ref{subsec:step1})}.} \mmi{Prophet executes target binaries with the simplified temporal prefetcher to collect counters through the user-space PMU interfaces, such as Linux's perf tools \cite{de2010new}. These collected counters are then analyzed in the subsequent \emph{Analysis} step to derive optimized metadata table management strategies.} \mmi{The \textit{simplified temporal prefetcher} operates with a configuration of Prophet with insertion policy disabled, a fixed metadata table of 1~MB, and a prefetching degree of 1.} This configuration ensures an unbiased evaluation of memory instructions under temporal prefetching, without incorporating any additional optimizations. \mmi{Compared to other profile-guided solutions \cite{khan2022whisper, song2022thermometer, jamilan2022apt, khan2021twig, khan2021ripple} that uses trace for profiling, Prophet is more \textbf{\emph{lightweight}}, introducing negligible profiling overhead in the Profiling step, as well as analysis and instruction overhead in the Analysis step (Section~\ref{subsec:evalight}).}

\textbf{Step2: Analysis \mmi{(Section~\ref{subsec:step2}).}} Prophet processes the counters collected in Step 1 through offline scripts, generating two types of hints \mmi{for efficient metadata table management: PC-level and application-level. PC-level hints are specific to individual memory instructions and are utilized in Prophet's insertion policy and replacement policy. On the other hand, application-level hints are applied globally through a CSR manipulation instruction at the program's start and are used in Prophet's resizing operations.} These hints are injected into the original binary, resulting in an optimized binary that can execute with Prophet.

\textbf{Step3: Learning \mmi{(Section~\ref{subsec:step3}).}} At regular intervals, Prophet samples new counters from varying inputs with the simplified Prophet and integrates them with previously collected counters. \mmi{Then, the subsequent Analysis step will generate new hints based on the merged counters, making Prophet's insertion policy, replacement policy, and resizing operations \textbf{\emph{adaptable}} to all encountered program inputs. Through repeated learning, Prophet innovatively enables a single optimized binary to achieve optimal performance across a wide range of program inputs.}

\section{Design}

\subsection{Step 1: Profiling}
\label{subsec:step1}

\begin{figure}[t]
\centering
\includegraphics[width=.99\linewidth]{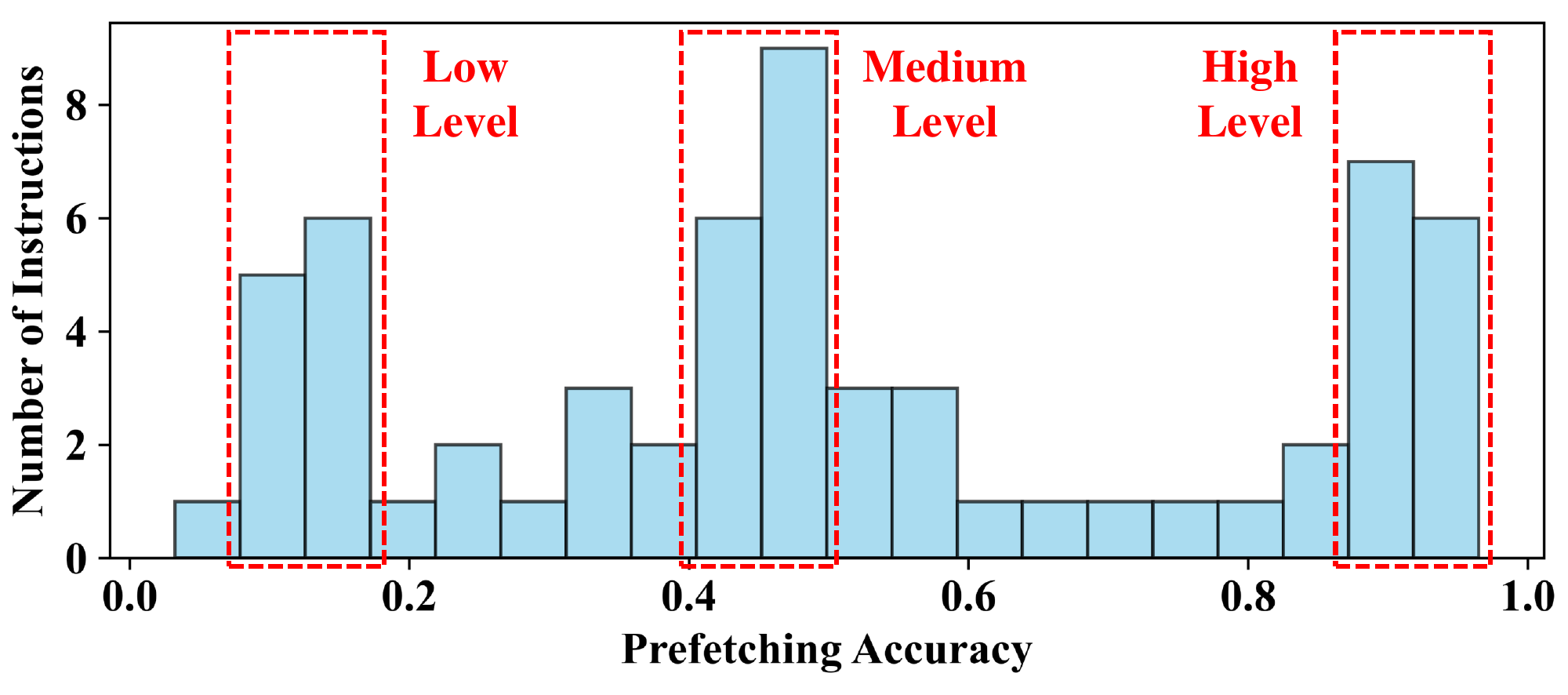}
\vspace{-.1in}
\caption{\mmi{The prefetching accuracy of temporal prefetching across different memory instructions in omnetpp.}}
\vspace{-.2in}
\label{fig:cdf}
\end{figure}

Two key questions arise in this step: \textit{What information is necessary to efficiently guide the management of the metadata table? How can we acquire this information with current architectures?} To answer the first question, we identify two primary objectives for managing the metadata table: (1) enhancing its utilization, and (2) reducing its impact on \mmi{the} LLC. The \emph{insertion} and \emph{replacement} policies are well-suited for achieving the first goal, while \emph{resizing} operations are suitable for the second. For enhancing utilization, \textbf{prefetching accuracy per memory instruction} serves as a critical metric. \mmi{As shown in Figure~\ref{fig:cdf}, although individual metadata accesses (Figure~\ref{fig:motiv}) exhibit high variability, the temporal prefetching accuracy of every instruction can be broadly classified into distinct levels. Since the prefetching accuracy reflects the adaptability of memory instructions to the temporal prefetcher, lower-level memory instructions generate fewer memory accesses exhibiting temporal patterns compared to higher-level memory instructions. Thus, the insertion policy can filter out metadata entries from lowest-level memory instructions, while the replacement policy can assign fine-grained replacement priorities to unfiltered metadata entries based on their levels. }For reducing the metadata table's impact on \mmi{the} LLC, the \textbf{number of allocated entries} in the metadata table is a useful metric. Sampling this metric at the program execution's end allows us to determine the maximum metadata table size.

We employ Intel's Processor Event-Based Sampling (PEBS) \cite{pebs} utility to collect PC-level prefetching accuracy. PEBS records the program context (e.g., PC) when specific events occur. Prior to data collection, we configure the temporal prefetcher in its simplified mode (Section~\ref{subsec:process}) and disable all other L2 prefetchers. We then enable PEBS to sample following events:

\begin{itemize}
    \item \texttt{MEM\_LOAD\_RETIRED.L2\_Prefetch\_Issue} counts the number of issued prefetch requests.
    \item \texttt{MEM\_LOAD\_RETIRED.L2\_Prefetch\_Useful} tracks the number of prefetches hit by demand requests.
\end{itemize}

The above two events can be implemented with minor modifications to existing \texttt{MEM\_LOAD\_RETIRED.L2\_MISS} (L2 cache miss) event, which is already supported on Intel's Xeon Processor \cite{perfmon}. Each PC's prefetching accuracy can be computed as:

\[Prefetching\ Accuracy = \frac{L2\_Useful\_Prefetches}{L2\_Issued\_Prefetches}\]

The number of allocated entries in the metadata table, an application-level metric, can be measured using standard PMU counters. For instance, we can define two counters: one for the number of metadata table insertions and another for replacements. The number of allocated entries in the metadata table can be calculated as: 

\[Allocated\ Entries = Insertions - replacements\]


\subsection{Step 2: Analysis}
\label{subsec:step2}

This step analyzes the counters collected in Step 1 to generate PC-level and application-level hints for the insertion policy, replacement policy, and resizing operations. Although both the insertion and replacement policies are guided by prefetching accuracy, Prophet adopts different strategies. The insertion policy filters out memory instructions that clearly lack temporal patterns (those with extremely low prefetching accuracy), while the replacement policy applies more refined management to the remaining memory instructions. Our observations indicate that as long as a memory instruction's prefetching accuracy is not particularly low, at least some of its memory accesses exhibit a temporal pattern.

\textbf{Prophet Insertion Policy} uses Equation~\ref{eq:insertion} to decide whether to use demand requests from PCs to train the temporal prefetcher and insert the corresponding metadata. \mmi{We define $EL\_ACC$ as an extremely low threshold for the prefetching accuracy, showing that memory instructions almost exhibit no temporal pattern.}

\begin{equation}
\label{eq:insertion}
    I(acc) = \left\{
             \begin{array}{ll}
             1, &  acc \geq EL\_ACC\\
             0, &  acc < EL\_ACC
             \end{array}
\right.
\end{equation}

The above equation indicates that if a PC's accuracy under temporal prefetching falls below $EL\_ACC$, Prophet instructs the temporal prefetcher to discard all demand requests associated with that PC. This decision is encoded as a one-bit hint injected into the corresponding memory instructions. All demand requests generated by these instructions will carry the embedded hint. Upon reaching the prefetcher, a simple logic checks the hint to decide whether to discard the corresponding requests.

\textbf{Prophet Replacement Policy} assigns a priority level to each stored metadata entry, with lower levels prioritized for replacement. Like the Prophet insertion policy, these priority levels are initially embedded in memory access instructions. When inserting new metadata entries, Prophet records their associated priority levels into the Prophet Replacement State.
 
We apply Equation~\ref{eq:repl} to determine each memory instruction's priority level. The \mmi{$n$} is a parameter controlled by the designer.

\begin{equation}
\label{eq:repl}
\mmi{
R(acc) = \left\{
\begin{array}{ll}
0, & EL\_ACC \leq acc < \frac{1}{2^n} \\
1, & \frac{1}{2^n} \leq acc < \frac{2}{2^n} \\
2, & \frac{2}{2^n} \leq acc < \frac{3}{2^n} \\
\vdots & \vdots \\
2^n-1, & \frac{2^n - 1}{2^n} \leq acc < 1
\end{array}
\right.
}
\end{equation}

When choosing victim entries in the metadata table, Prophet first identifies victim candidates with the lowest priority level. Then, Prophet applies LRU among these victim candidates to determine the final replaced entry.

\textbf{Prophet Resizing} estimates the metadata table size based on the number of allocated metadata entries at the end of program execution. Assuming the counter value is $S$, we first round it\footnote{we ensure the rounded value does not exceed the maximum number of entries that a 1MB metadata table can accommodate.} to the nearest power of 2, and then use the Equation~\ref{eq:resizing} to determine the number of ways allocated for the metadata table in the LLC:

\begin{equation}
\label{eq:resizing}
    Allocated\ Ways = ceil \left( \frac{Target\ Size}{Number\ of\ sets\ in\ LLC} \right)
\end{equation}

\mmi{We completely disable temporal prefetching when the outcome of the above equation is less than 0.5.} Based on the result of Equation~\ref{eq:resizing}, Prophet inserts a CSR manipulation instruction at the program's start to configure the metadata table size.

\subsection{Step 3: Learning}
\label{subsec:step3}

\begin{figure}[t]
\centering
\includegraphics[width=.98\linewidth]{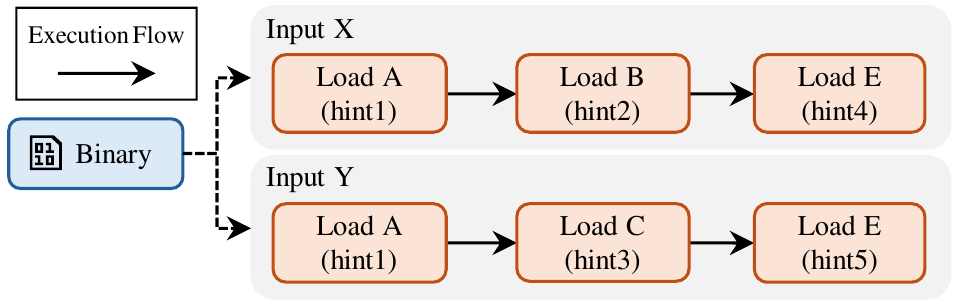}
\vspace{-.1in}
\caption{Challenges of traditional profile-guided optimizations across different program inputs.}
\vspace{-.2in}
\label{fig:adjust}
\end{figure}

In this step, Prophet collects counters from varying program inputs and integrates them, allowing the optimized binary to deliver optimal performance across diverse program inputs. Figure~\ref{fig:adjust} illustrates \textit{why profile-guided methods are sensitive to different program inputs} and \textit{why the hints derived from one input may not be applicable to other inputs}. A single binary may follow different execution paths depending on the input. For example, with input X, the binary might execute memory access instructions A, B, E, while input Y causes it to execute instructions A, C, E. When applying profile-guided optimizations to this binary under inputs X and Y respectively, there are three different scenarios:

\begin{itemize}
    \item \textbf{Load A}: Under both inputs X and Y, the binary executes Load A. Profile-guided optimizations generate identical hints for both inputs, possibly because they execute the same code and produce similar profiling metrics. In this case, the hints generated for Load A under input X remain effective for Load A under input Y.
    \item \textbf{Loads B and C}: execute completely different instructions, resulting in distinct hint information. In this case, any hint derived from Load B under input X is ineffective for Load C under input Y. 
    \item \textbf{Load E}: Although both inputs X and Y execute Load E, the global execution context impacts Load E differently under each input. Consequently, profile-guided methods may generate distinct hints. In this case, the hints generated for Load E under input X are ineffective for Load E under input Y.
\end{itemize}

Building on the three cases above, we develop a process that enables our counter-based profile-guided optimizations to adapt to different program inputs. Prophet maintains the counters from step 2 when it progresses to step 3 (e.g., input X). In step 3, we assume Prophet acquires new counters under previously unseen inputs (e.g., input Y). Prophet merges them with the previously maintained counters. For prefetching accuracy per PC, we use the Equation~\ref{eq:mergepc} for merging. The variable $o$ indicates old counter value under input X, while $n$ represents new counter value under input Y. The variable $l$ indicates the number of Prophet loops, where each execution of step 2 counts as one loop, and $L$ is a parameter predefined by the designer.

\begin{equation}
\label{eq:mergepc}
    Merged = \left\{
    \begin{array}{lr}
    o + \frac{1}{\min(l + 1, L)} \times (n - o), & \exists o \in X  \\
    n,    &  \nexists o \in X
    \end{array}
\right.
\end{equation}

For the number of allocated metadata entries at the end of
program execution, we apply the Equation~\ref{eq:maxentries} for merging:

\begin{equation}
\label{eq:maxentries}
    Merged = \max(o, n)
\end{equation}

We assume a binary first encounters input X (steps 1 and 2), followed by input Y (steps 3 and 2). Next, we will prove that the optimized binary can ultimately adapt to both the previous and the newly observed inputs.

\begin{itemize}
    \item \textbf{Merged prefetching accuracy for Load A}: Since Load A could receive the same hints under input X and Y, both $o$ and $n$ fall within the same range as defined by Equation~\ref{eq:insertion} and Equation~\ref{eq:repl}. After applying Equation~\ref{eq:mergepc} with $l$ = 1, the merged accuracy remains within the same range, resulting in the same hint being generated in the next step 2.
    \item \textbf{Merged prefetching accuracy for Loads B and C}: Prior to input Y, Prophet lacks counters for Load C (i.e., $\nexists o \in X$). Therefore, the merged prefetching accuracy for Load C is set to $n$. In the subsequent Step 2, Prophet injects new hints for Load C based on $n$. As a result, Prophet successfully learns hints for Load C, which was previously unrecorded.
    \item \textbf{Merged prefetching accuracy for Load E}: Load E could receive different hints under input X and Y, causing $o$ and $n$ to fall into different ranges in Equation~\ref{eq:insertion} or Equation~\ref{eq:repl}. In this case, Equation~\ref{eq:mergepc} adjusts $o$ through the offset between $n$ and $o$. If Prophet observes higher prefetching accuracy for Load E under input Y ($n - o > 0$), it increases the estimated accuracy, refining its hints accordingly. Conversely, if $n - o < 0$, Prophet decreases the estimated accuracy. Over time, frequently observed counter values dominate merged results.
    \item \textbf{Merged allocated entries for the entire program}: Prophet adopts a conservative strategy to accommodate the metadata table size requirements for all program inputs.
\end{itemize}


\subsection{Hint Information Injection}
\label{subsec:hint}

According to Equation~\ref{eq:insertion} and Equation~\ref{eq:repl}, each memory instruction requires at most 3-bit hint information. We design two methods for injecting this 3-bit information into memory access instructions.

\textbf{Hint buffer.} Reference to the approach in Whisper \cite{khan2022whisper}, we can leverage specialized hint instructions to carry hints. When these hint instructions are executed, Prophet stores the hint information and PC tag in a hint buffer near the temporal prefetcher. \mmi{Hint instructions are only required to execute once. To minimize their impact on total dynamic instructions, they can be inserted at the entry point of programs using BOLT \cite{panchenko2019bolt}.} Prophet applies its optimizations to memory instructions whose PC matches an entry in the hint buffer. To efficiently utilize the hint buffer, Prophet focuses on memory instructions that contribute the most to cache misses. We pinpoint these instructions by using \texttt{MEM\_LOAD\_RETIRED.L2\_MISS} event at step 1. Empirical findings show that a 128-entry hint buffer (0.19~KB) is sufficient for achieving high performance. Although this approach introduces additional storage overhead, it is compatible with all instruction set architectures.

\textbf{Reserved bits \mmi{or instruction prefix}.} We can embed hints into reserved bits within instructions, allowing hints to be decoded and combined with memory access instructions. This approach saves additional space for storing hints but is constrained by the requirement that commonly used memory access instructions include reserved bits, limiting its applicability. \mmi{For CISC architectures like x86, we can also add prefixes for memory instructions to carry hints. While this method does not require reserved bits, it increases the code footprint and may impact I-cache performance. However, Prophet focuses on at most 128 memory instructions and introduces only a 3-bit prefix for these instructions. Therefore, Prophet maximally introduces $\frac{3 \times 128}{64} = 6$ Byte storage overhead to I-cache (usually with 64~KB). Consequently, the x86 instruction prefix scheme has an almost negligible impact on I-cache performance.}

\subsection{Multi-path Victim Buffer}
\label{subsec:reuse}

\begin{figure}[t]
\centering
\includegraphics[width=.98\linewidth]{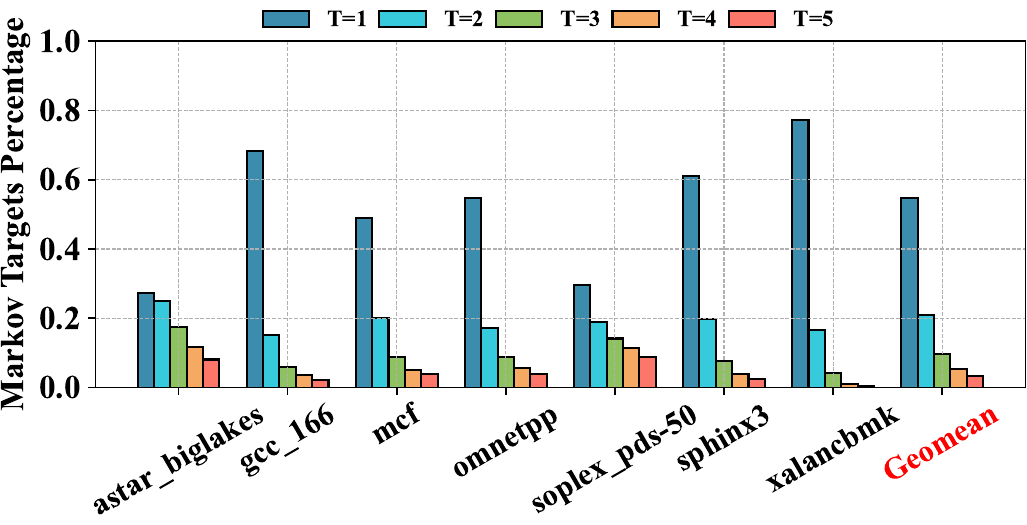}
\vspace{-.15in}
\caption{\mmi{The percentage of Markov target number (T) in temporal prefetching.}}
\vspace{-.2in}
\label{fig:suc}
\end{figure}

We observe that the same memory address can appear in multiple distinct temporal patterns. For example, if memory access sequences (A, B, C) and (A, B, D) exhibit temporal patterns, B has two potential Markov targets: C and D. \mmi{As shown in Figure~\ref{fig:suc}, 54.85\%, and 20.88\%, 9.71\% of memory addresses in the SPEC CPU benchmark have 1, 2, and 3 Markov targets, respectively.} However, previous on-chip temporal prefetchers \cite{ainsworth2024triangel, wu2019temporal, wu2021practical} store only one target per Markov entry, often resulting in inaccurate prefetches and unsolvable demand accesses. To address this issue, simply storing multiple prefetching candidates in the metadata table is impractical, as it significantly increases storage overhead.

\mmi{In order to efficiently handle the above scenario, Prophet introduces a Multi-path Victim Buffer, allowing it to store Markov targets that have been evicted from the metadata table. Prophet manages Multi-path Victim Buffer based on the following rules:}


\begin{figure}[t]
\centering
\includegraphics[width=.98\linewidth]{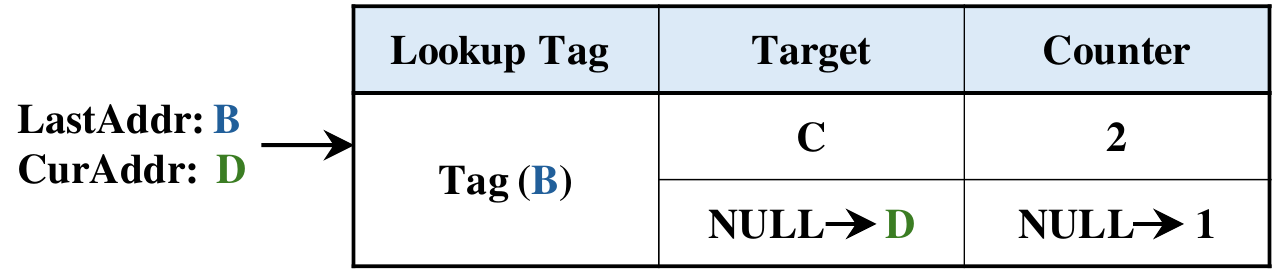}
\vspace{-.1in}
\caption{The Multi-path Victim Buffer.}
\vspace{-.2in}
\label{fig:reuse}
\end{figure}

\begin{itemize}
    \item \mmi{\textbf{Insertion.}
    To efficiently utilize the space of Multi-path Victim Buffer, we store only Markov targets whose priority levels (Equation~\ref{eq:repl}) are greater than 0 ($acc > EL\_ACC$).}
    \item \mmi{\textbf{Replacement.} We reuse Prophet Replacement Policy to maintain frequently used Markov targets. As shown in Figure~\ref{fig:reuse}, we add a counter for each Markov target, incrementing the counter value each time the target is accessed. We set the priority levels as the maximal\footnote{If we buffer two or more Markov targets per metadata entries} counter value of Markov targets (which differs from Equation~\ref{eq:repl}).}
    \item \mmi{\textbf{Prefetch.} Prefetching with the Reuse Buffer or metadata table will also trigger a lookup in the Multi-path Victim Buffer. Prophet uses the same lookup addresses to search for entries in the Multi-path Victim Buffer. If Prophet detects different Markov targets, it prefetches these targets accordingly.} 
\end{itemize}
\section{Evaluation}
\label{sec:eva}

\subsection{Experimental Setup}
\label{sec:expsetup}

\begin{table}[h]
\centering
\vspace{-3.5mm}
\caption{System Configuration.}
\vspace{-2mm}
\label{tab:config}
\begin{tabular}{|l|l|}
\hline
\textbf{Module} & \textbf{Configuration} \\
\hline
\hline 
Core &5-\mmi{wide} fetch, 5-\mmi{wide} decode \\
&10-\mmi{wide} issue, 10-\mmi{wide} commit \\
&120-entry IQ, 85/90-entry LQ/SQ \\
& 288-entry ROB\\
\hline
Private L1 I/D cache & 64~KB each, 4-way, 64B line, 16 MSHRs  \\
&PLRU, 2 cycles hit latency \\
&degree-8 stride prefetcher for L1D cache \\
\hline
Private L2 cache & 512~KB, 8-way, 64B line, 32 MSHRs \\
&PLRU, mostly\_inclusive \\
&9 cycles hit latency \\
\hline
Shared L3 cache & 2~MB/core, 16-way, 64B line, 36 MSHRs \\
&CHAR \cite{chaudhuri2012introducing}, mostly\_exclusive \\
&20 cycles hit latency \\
\hline
Memory 
& LPDDR5\_5500\_1x16\_BG\_BL32 \\
& Single channel, 1 rank per channel \\
\hline
\end{tabular}
\vspace{1mm}
\vspace{-4mm}
\end{table}

\textbf{System Configuration.} We evaluate Prophet using gem5's FS mode \cite{binkert2011gem5}. We utilize facilities within gem5 to collect counters required by Prophet (Section~\ref{subsec:step1}). Following the rules defined in Section~\ref{subsec:step2}, we use an offline script to analyze these counters and generate hints, which are then injected into binaries via the hint buffer (Section~\ref{subsec:hint}). Our simulation environment adopts parameters almost consistent with those utilized in the Triangel \cite{ainsworth2024triangel}. The primary system configurations are outlined in Table~\ref{tab:config}. \mmi{Prophet and Triangel are trained on the L2 cache access stream, including prefetch requests generated by L1 stride prefetchers.} 



\textbf{Workloads.} Following previous temporal prefetchers \cite{wu2019efficient,wu2019temporal,wu2021practical,ainsworth2024triangel} and software indirect prefetching schemes \cite{zhang2024rpg2, jamilan2022apt, ainsworth2017software, khan2021dmon}, we evaluate Prophet with irregular SPEC CPU workloads and graphic workloads (i.e., CRONO \cite{ahmad2015crono}). These workloads exhibit diverse memory access patterns that are representative of a wide range of benchmarks. We apply the SimPoint technique \cite{sherwood2002automatically} to generate checkpoints across all workloads. Each \yao{SimPoint-sampled} checkpoint is warmed up with 250M instructions, followed by a simulation of the next 50M instructions. The reported performance metrics for each benchmark are calculated by aggregating the results from all its checkpoints with weighted averages.


\textbf{Baseline.} We compare Prophet against the state-of-the-art hardware temporal prefetcher, Triangel \cite{ainsworth2024triangel}, and software indirect indirect prefetching scheme, RPG$^2$ \cite{zhang2024rpg2}. For Triangel, we utilize the open-source implementation provided by its original paper \cite{triangel}, preserving its complete functionality. \mmi{For RPG$^2$, we follow its original methodology, first identifying memory instructions that result in at least 10\% cache misses and have prefetch kernels supported by RPG$^2$. Then, we utilize the hint buffer mechanism (Section~\ref{subsec:hint}) to simulate prefetch instruction insertion. Specifically, we record the PC of identified memory instructions along with an initial prefetch distance in the hint buffer. Upon encountering recorded PCs, we issue a prefetch request where the target address equals the accessed memory address + distance. Finally, we tune the distance using RPG$^2$'s binary search method and record the performance with the optimal distance as the final report.}

\subsection{Performance}
\label{subsec:perf}

\begin{figure}[t]
\centering
\includegraphics[width=.99\linewidth]{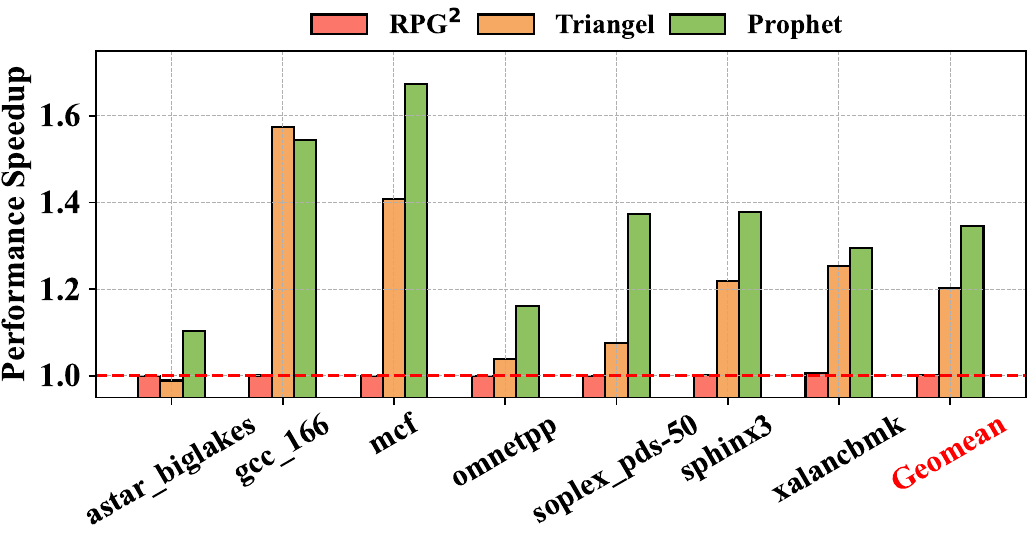}
\vspace{-.1in}
\caption{IPC speedup compared to RPG$^2$ and Triangel.}
\vspace{-.2in}
\label{fig:perf}
\end{figure}

\begin{figure}[t]
\centering
\includegraphics[width=.99\linewidth]{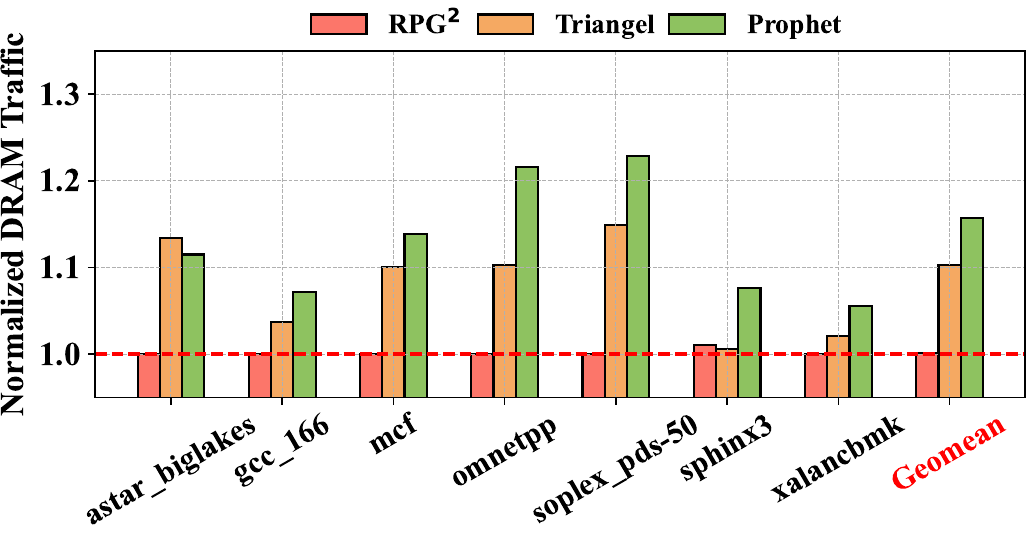}
\vspace{-.1in}
\caption{DRAM traffic compared to RPG$^2$ and Triangel.}
\vspace{-.2in}
\label{fig:dram}
\end{figure}

\begin{figure}[t]
\centering
\includegraphics[width=.99\linewidth]{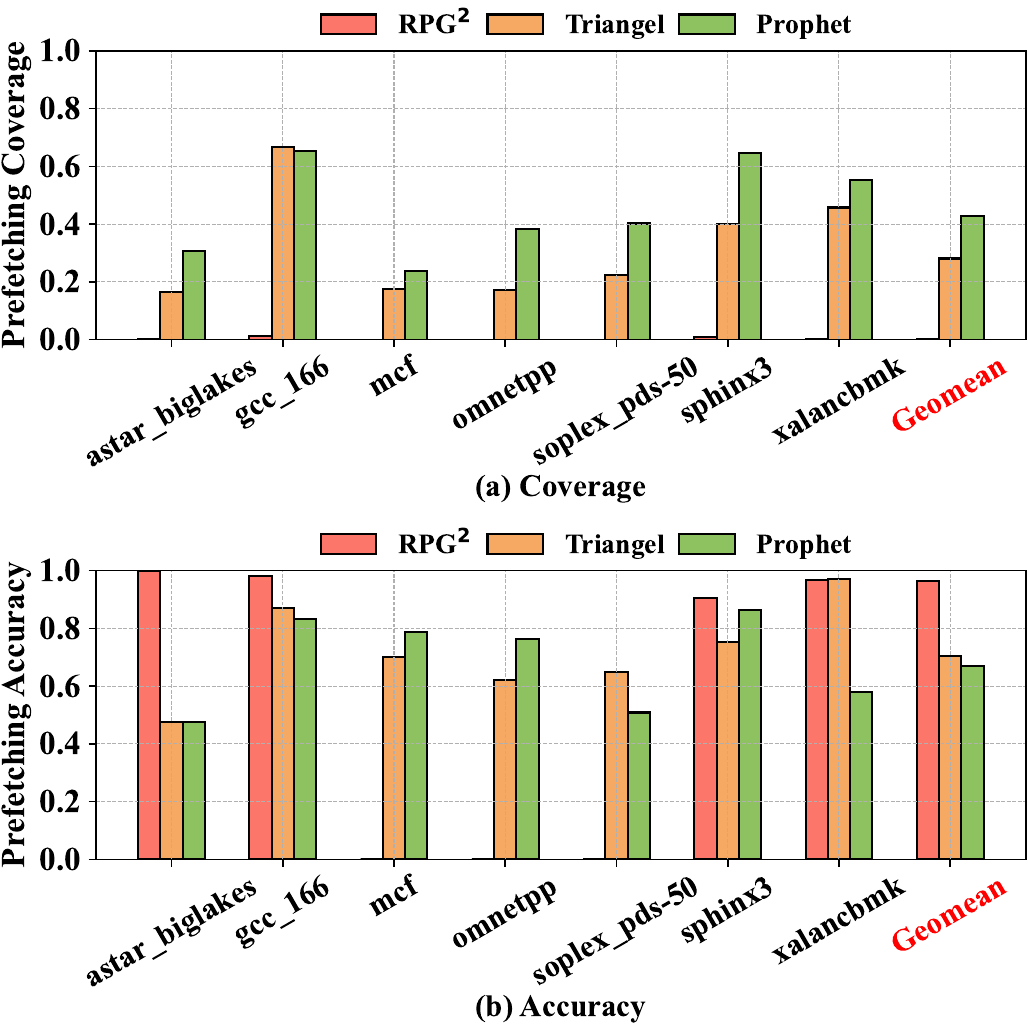}
\vspace{-.15in}
\caption{\mmi{Prefetching coverage and accuracy}\protect\footnotemark.}
\vspace{-.25in}
\label{fig:acccov}
\end{figure}

Figure~\ref{fig:perf} compares the performance of Prophet with RPG$^2$ and Triangel. The results show that Prophet achieves a \mmi{34.58\%} speedup over the baseline without temporal prefetchers, compared to 0.1\% for RPG$^2$ and 20.35\% for Triangel, outperforming RPG$^2$ by \mmi{34.48\%} and Triangel by \mmi{14.23\%}. Figure~\ref{fig:dram} shows that Prophet induces 18.67\% memory traffic \mmi{(cumulative DRAM reads + DRAM writes)}, compared to 0.07\% for RPG$^2$ and 10.33\% for Triangel, indicating that Prophet’s performance gain over Triangel results in only \mmi{5.35}\% additional memory traffic. The workload-specific performance improvements are comparable to those reported in the original Triangel paper \cite{ainsworth2024triangel}. However, the overall speedup for Triangel in our experiments is not identical because we use SimPoint to generate checkpoints instead of the original method described in \cite{triangel}, which evenly samples checkpoints throughout the program's lifecycle, potentially misrepresenting actual program execution.

\footnotetext{\mmi{RPG$^2$ does not identify qualified prefetch kernels for mcf, omnetpp, and soplex, so we set their prefetching accuracy to 0.}}

\textbf{Analysis: Prophet versus RPG$^2$.} Our experimental results demonstrate that prior profile-guided indirect prefetching schemes are ineffective for most temporal patterns. We observe that most active memory access instructions (i.e., those causing $>$90\% cache misses) in the evaluated workloads exhibit pointer-chasing patterns or indirect access patterns where the prefetch kernel does not follow stride patterns. As described in Section~\ref{subsec:drawback_soft}, prior solutions cannot handle these cases, leading to limited performance improvements.

\textbf{Analysis: Prophet versus Triangel.} Prophet consistently outperforms Triangel across most workloads. To validate this, we analyze the temporal prefetcher’s prefetching coverage and accuracy, as shown in Figure~\ref{fig:acccov}(a) and Figure~\ref{fig:acccov}(b). Prophet reduces demand misses by \mmi{42.75\%}, compared to 28.08\% for Triangel. For prefetching accuracy, Prophet performs comparably to Triangel, indicating that Prophet's performance gain comes from more efficient metadata storage management rather than aggressive prefetching. For example, Prophet can simultaneously enhance both prefetching coverage and accuracy in workloads such as mcf and omnetpp. In the case of gcc, which is particularly sensitive to cache pollution, Prophet’s performance gain is slightly lower than Triangel.

\subsection{Adaptable: Different Program Inputs}

\begin{figure}[h]
\centering
\includegraphics[width=.99\linewidth]{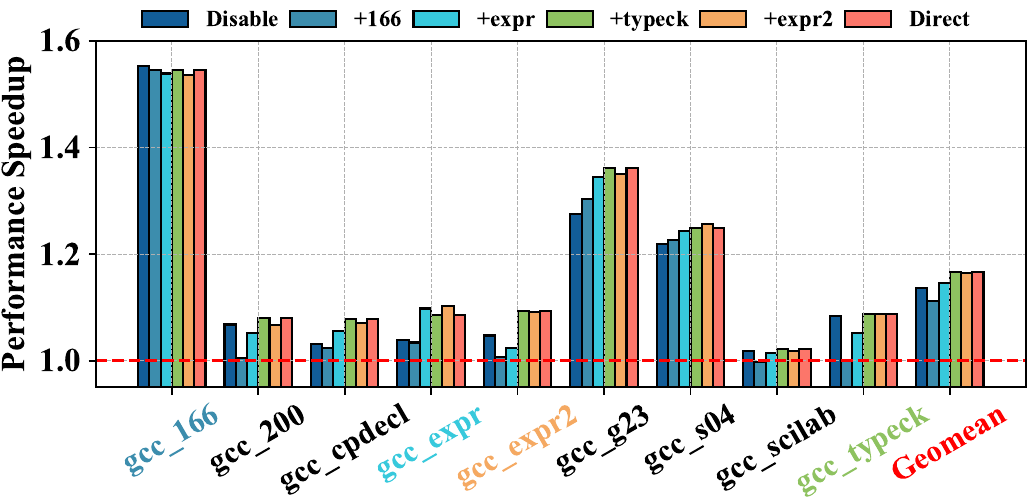}
\vspace{-.1in}
\caption{Prophet learns counters from gcc's inputs.}
\vspace{-.2in}
\label{fig:input}
\end{figure}

\begin{figure}[h]
\vspace{-.1in}
\centering
\includegraphics[width=.99\linewidth]{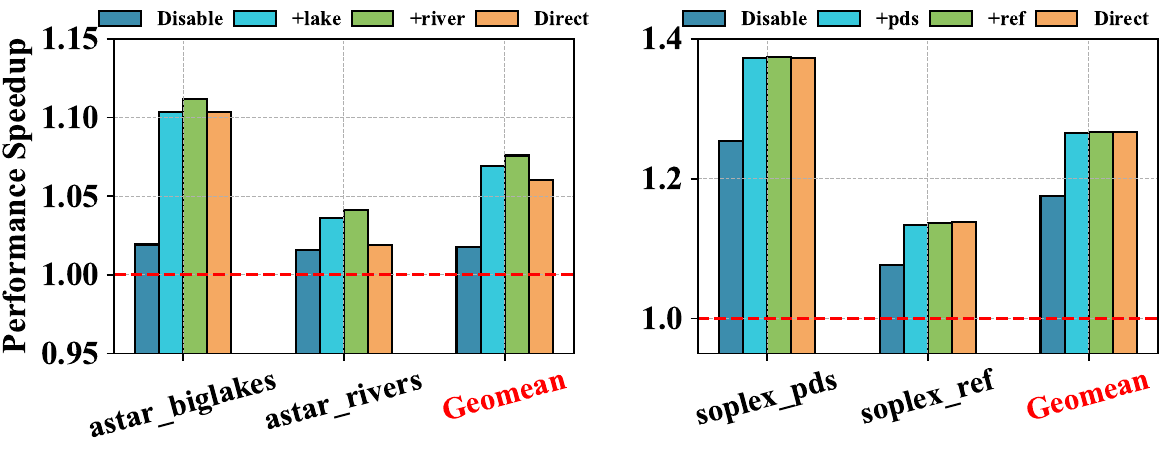}
\vspace{-.1in}
\caption{\mmi{Prophet's learning feature can be generalized to other workloads, such as astar and soplex.}}
\vspace{-.2in}
\label{fig:input2}
\end{figure}

Figure~\ref{fig:input} evaluates Prophet's adaptability by making it iteratively learn counters from different program inputs within a single application. \mmi{The leftmost bar shows the performance of Triage4 + Triangel metadata (Section~\ref{subsec:breakdown}), indicating the status where no input is fed to Prophet (shown as ``Disable'').} Then, we provide the inputs in the following order: gcc\_166, gcc\_expr, gcc\_typeck, and gcc\_expr2. The first input, gcc\_166, is processed in Step 1, while the subsequent inputs are integrated in Step 3, as shown in Figure~\ref{fig:process_overview}. \mmi{To evaluate Prophet's ``ideal'' performance for each input, we directly profile these inputs and make Prophet learn their respective counters (learning goal, shown in the rightmost bar as ``Direct'').}

Figure~\ref{fig:input} demonstrates that, through repeated learning, Prophet enables a single optimized binary to achieve optimal performance across various program inputs. Initially, when learning only from gcc\_166, Prophet delivers sub-optimal performance on gcc\_expr, gcc\_typeck, and gcc\_expr2. However, as Prophet incorporates additional inputs, it progressively achieves optimal performance across all of them. Notably, even without directly learning from inputs like gcc\_200, Prophet improves its performance due to counters learned from other inputs, such as gcc\_expr, which share similar memory access patterns. These results indicate that Prophet can achieve optimal performance across all program inputs with fewer training iterations than the total number of inputs. Consequently, with only 4 rounds of learning, Prophet achieves near-optimal performance across all 9 gcc inputs. \mmi{Furthermore, Figure~\ref{fig:input2} demonstrates the learning features in Prophet can be generalized to other workloads.}

\textbf{Prophet versus Other profile-guided solutions.} To the best of our knowledge, existing profile-guided solutions lack an adaptive mechanism for learning counters or traces from different inputs. These solutions either deliver sub-optimal performance when encountering new inputs or require fresh profiling (e.g., RPG$^2$) but cannot leverage prior profiling data, making them unable to adapt to previously encountered inputs. In contrast, Prophet’s adaptability makes it more practical for deployment in commercial processors.

\subsection{Lightweight: Profiling, Analysis, Instruction Overhead}
\label{subsec:evalight}

\subsubsection{Profiling Overhead} 

Prophet's profiling overhead arises from collecting counters during Step 1 and Step 3 in Figure~\ref{fig:process_overview}. Prophet utilizes the PEBS and standard PMU to gather these counters (Section~\ref{subsec:step1}), so the profiling overhead depends on the implementation of PEBS and PMU. According to \cite{bitzes2014overhead}, sampling 4 PEBS events incurs less than 2\% performance overhead, while sampling a single standard PMU event incurs negligible overhead. Given Prophet only requires sampling two or three PEBS events (depending on the ways to inject hints) and one standard PMU event, it incurs less than 2\% profiling overhead. Importantly, not every program execution requires profiling. Prophet only sample counters at intervals (Section~\ref{subsec:process}), determined by the complexity of programs and input variety. Empirically, profiling once every 10–100 executions suffices. Most program executions incur no profiling overhead. Moreover, we can stop profiling when further performance gains are minimal.


\subsubsection{Analysis Overhead} 

Prophet's analysis overhead comes from analyzing the collected counters to generate hints during Step 2 in Figure~\ref{fig:process_overview}. Across all evaluated workloads, Prophet's analysis overhead is negligible, less than one second. Furthermore, as with profiling, not every program execution incurs analysis overhead; only those executions in which Prophet is enabled to collect counters will require analysis, further reducing the overall analysis overhead.

\subsubsection{Instruction Overhead} 
\label{sec:instoverhead}
Prophet's instruction overhead depends on the hint injection methods described in Section~\ref{subsec:hint}. If reserved bits within instructions are used to carry hints, Prophet incurs no additional instruction overhead. In contrast, if specialized hint instructions are employed, the instruction overhead corresponds to the number of inserted hint instructions. In our setup, we insert a maximum of 128 hint instructions \mmi{at the entry point of programs (Section~\ref{subsec:hint})}. Compared to evaluated SPEC CPU workloads containing billions of instructions \cite{nair2008simulation}, Prophet introduces almost negligible overhead \mmi{for overall static and dynamic instructions}. 


\subsection{Generalization: Graphic Workloads}

\begin{figure}[t]
\centering
\includegraphics[width=.99\linewidth]{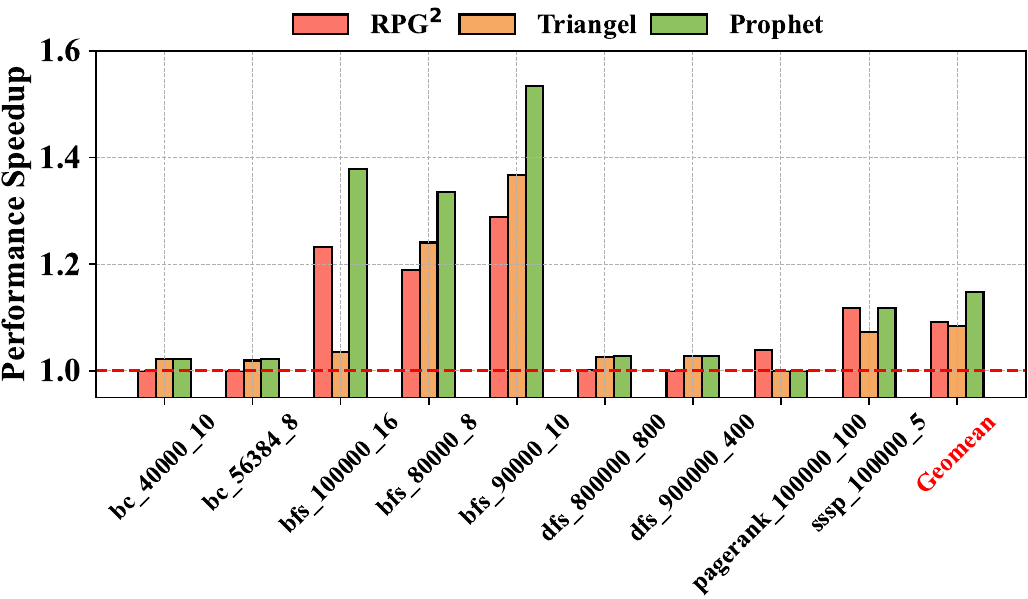}
\vspace{-.1in}
\caption{IPC speedup on graph workloads.}
\vspace{-.2in}
\label{fig:graph}
\end{figure}

Figure~\ref{fig:graph} evaluates Prophet's performance using CRONO \cite{ahmad2015crono}, a benchmark suite widely employed in indirect access prefetching schemes \cite{zhang2024rpg2, jamilan2022apt, ainsworth2017software, khan2021dmon}. Experimental results demonstrate that Prophet provides a performance speedup of \mmi{14.85\%} over the baseline with a hardware stride prefetcher alone, compared to 9.11\% for RPG$^2$ and 8.41\% for Triangel. Unlike SPEC CPU benchmarks, CRONO features more prefetch kernels with stride patterns, aligning with RPG$^2$'s strengths. As a result, RPG$^2$ delivers greater performance gains on CRONO. Prophet outperforms RPG$^2$ by handling more complex temporal patterns beyond RPG$^2$'s scope.

\subsection{\mmi{Sensitivity: Parameters in Prophet}}
\label{sec:sensitivity}

\begin{figure}[t]
\centering
\includegraphics[width=.99\linewidth]{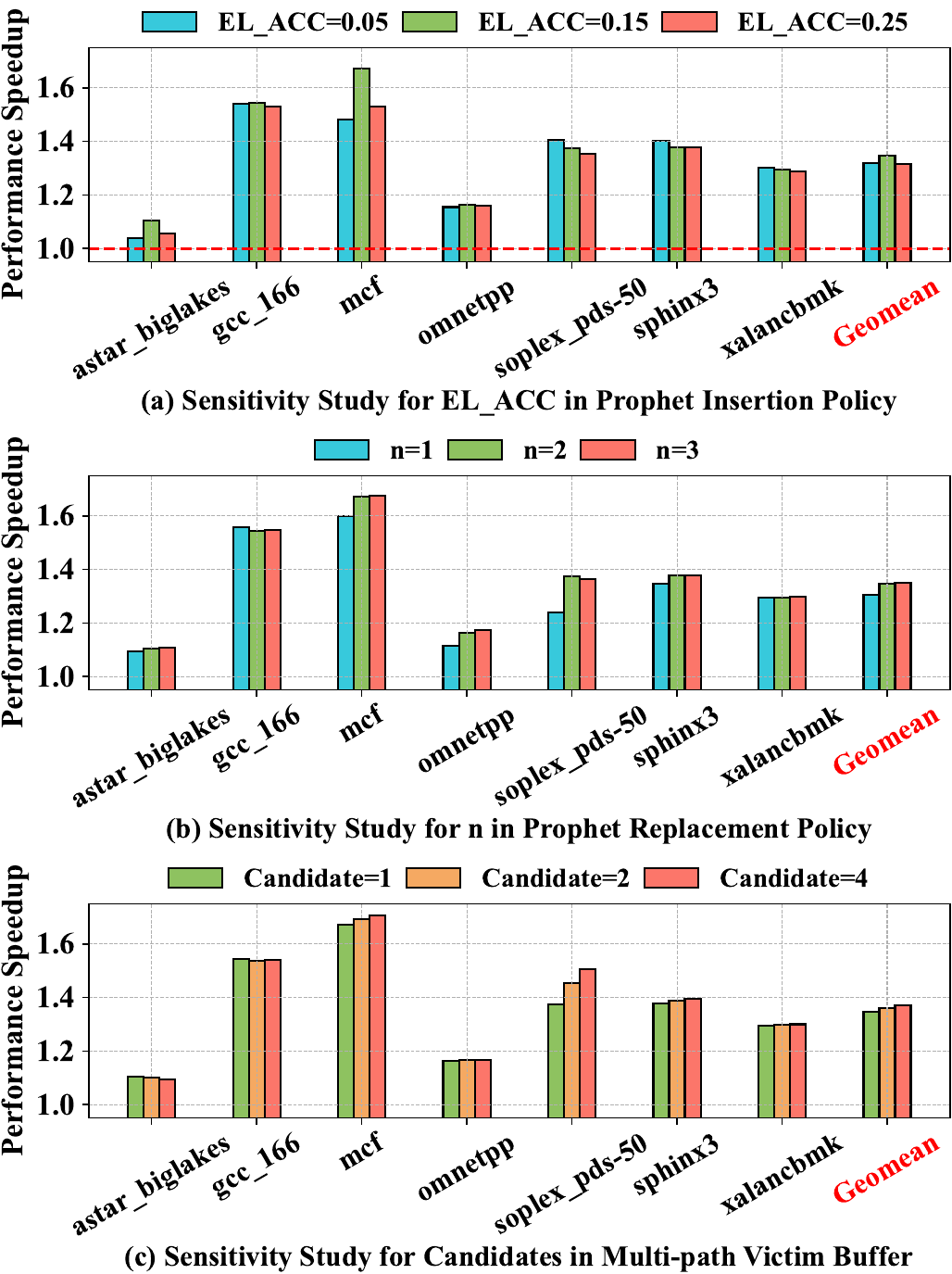}
\vspace{-.1in}
\caption{\mmi{Sensitivity study.}}
\vspace{-.22in}
\label{fig:sen}
\end{figure}

\mmi{Figure~\ref{fig:sen} evaluates primary parameters\footnote{\mmi{The green bar represents the parameters used in other experiments.}} in Prophet, such as $EL\_ACC$ for the Prophet Insertion Policy, $n$ for the Prophet Replacement Policy, and the candidates per metadata entry for the Multi-path Victim Buffer.}

\mmi{For $EL\_ACC$, we observe that both too high and too low parameter values can negatively impact performance. A low $EL\_ACC$ causes the metadata table to buffer too many metadata entries that do not exhibit temporal patterns, while a high $EL\_ACC$ has the opposite effect, potentially filtering out valuable entries.}

\mmi{For $n$, we observe that introducing additional bits to enable fine-grained classification of temporal patterns results in improved performance. However, the performance improvement is limited, and this modification introduces additional storage overhead for the Prophet Replacement State. To balance the performance gain with the storage overhead, we adopt a moderate configuration (n = 2, 2-bit Prophet Replacement State).}

\mmi{For the number of Markov candidates in the Multi-path Victim Buffer, we observe that having one candidate per metadata entry achieves the best trade-off between performance gain and storage overhead. While prefetching more candidates improves prefetching coverage, it can negatively impact performance. For example, prefetching additional Markov targets causes performance slowdown in astar, which is sensitive to cache pollution and memory bandwidth wastage. Furthermore, as shown in Figure~\ref{fig:suc}, a large proportion of memory addresses tend to have fewer than two Markov targets, resulting in unnecessary storage overhead when excessive Markov candidates are reserved in the buffer.}

\subsection{Sensitivity: Impact of L1 Prefetcher}

\begin{figure}[t]
\centering
\includegraphics[width=.99\linewidth]{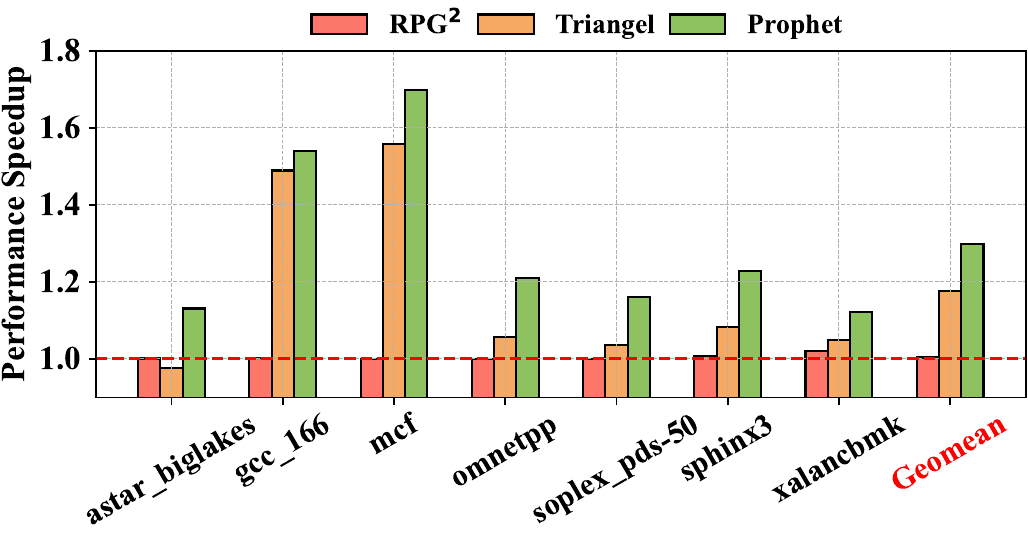}
\vspace{-.1in}
\caption{IPC speedup with varying L1 prefetcher.}
\vspace{-.25in}
\label{fig:ipcp}
\end{figure}

Figure~\ref{fig:ipcp} evaluates Prophet's performance with varying L1 prefetchers. As outlined in Table~\ref{tab:config}, our system configuration aligns with Triangel's setup for consistent comparison. However, commercial processors typically include additional L1 prefetchers since they can leverage more information (e.g., virtual addresses) and prefetch across page boundaries. For example, Arm’s Neoverse V2 \cite{arm-neov2} integrates stream, stride, and spatial prefetchers in the L1 cache. To assess Prophet’s performance with a more realistic L1 prefetcher configuration, we replace the L1 stride prefetcher with IPCP \cite{pakalapati2020bouquet}, simulating the setup in Arm’s Neoverse V2.

Figure~\ref{fig:ipcp} demonstrates that Prophet outperforms both RPG$^2$ and Triangel, achieving a performance speedup of \mmi{29.95\%} over the baseline without a temporal prefetcher. In comparison, RPG$^2$ shows a modest speedup of 0.36\%, while Triangel achieves a speedup of 17.51\%. Prophet consistently outperforms other schemes across all evaluated workloads. These results demonstrate that Prophet's performance improvement can also be applied to more complex L1 prefetcher configurations.

\subsection{Sensitivity: Memory Bandwidth}

\begin{figure}[h]
\vspace{-.15in}
\centering
\includegraphics[width=.99\linewidth]{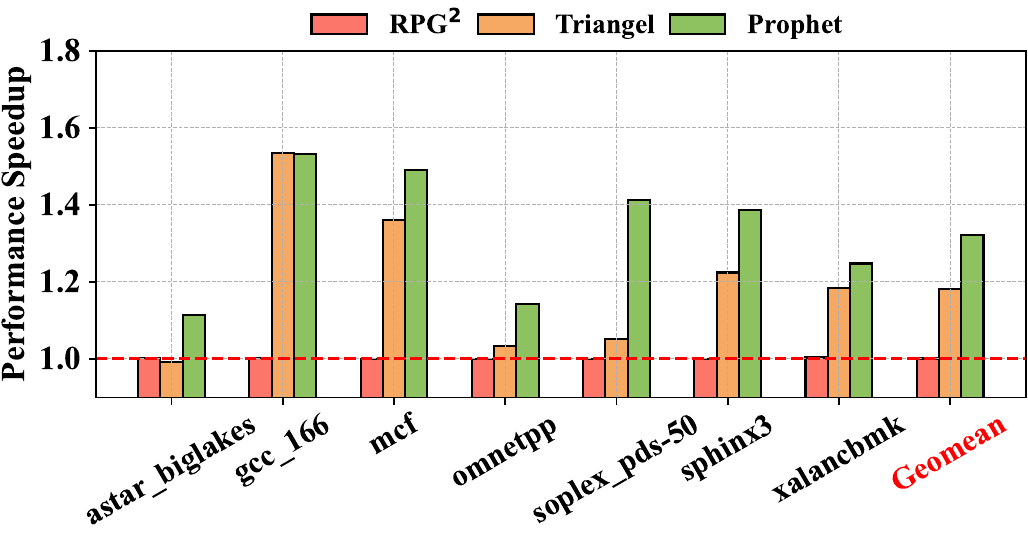}
\vspace{-.1in}
\caption{IPC speedup with varying DRAM channels.}
\label{fig:twochan}
\end{figure}

Figure~\ref{fig:twochan} evaluates Prophet, RPG$^2$, and Triangel with an increased number of DRAM channels. Prophet provides a performance speedup of \mmi{32.27\%} over the baseline without temporal prefetchers, compared to 0.1\% for RPG$^2$ and 18.17\% for Triangel. The experimental results show that Prophet remains effective across varying memory bandwidth environments.


\subsection{Prophet Features Breakdown}
\label{subsec:breakdown}

\begin{figure}[t]
\centering
\includegraphics[width=.99\linewidth]{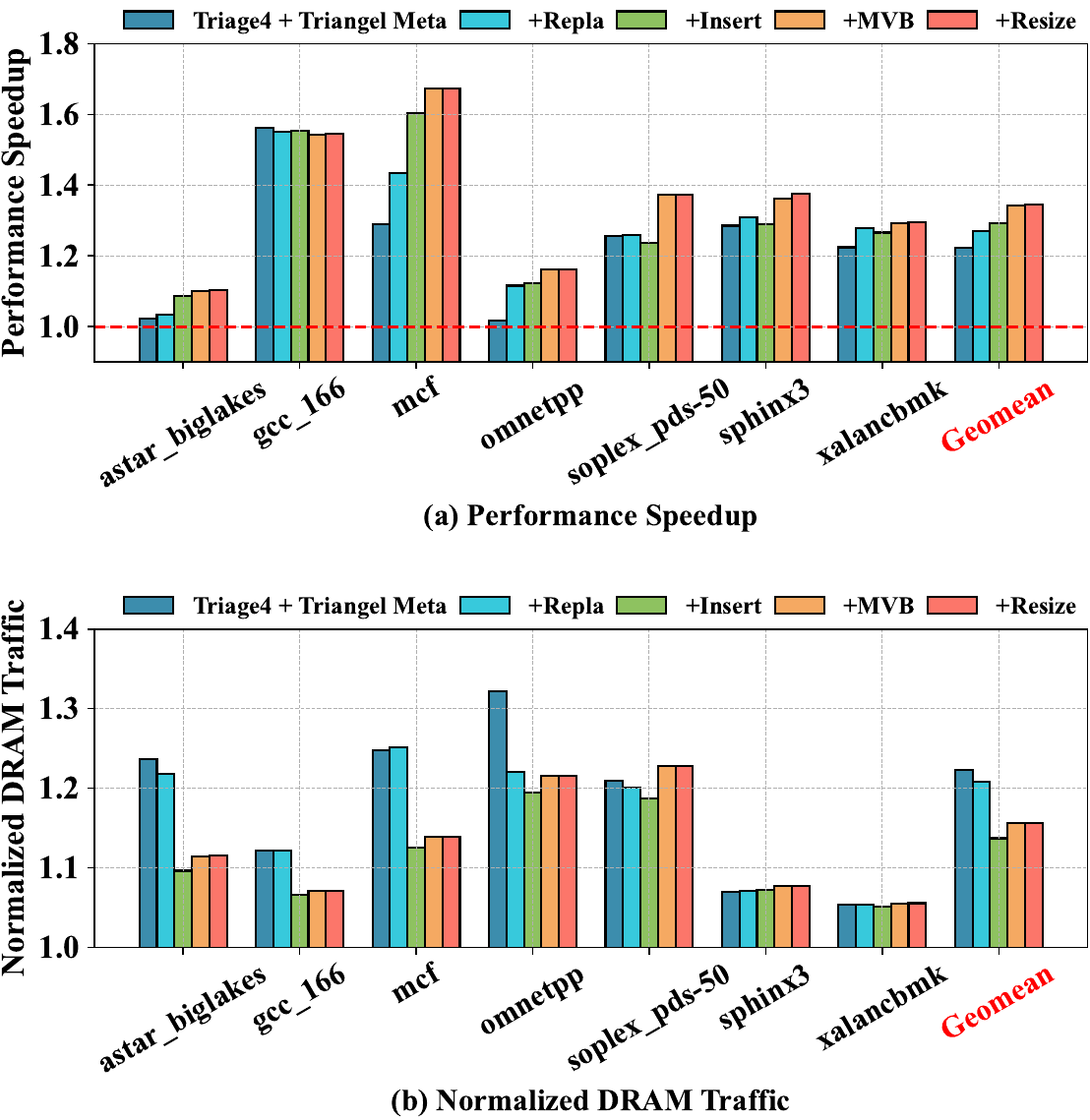}
\vspace{-.1in}
\caption{Prophet Features Breakdown.}
\vspace{-.15in}
\label{fig:ablation}
\end{figure}

Figure~\ref{fig:ablation} illustrates the contribution of each feature provided by Prophet. Our ablation study begins with Triage at a prefetch degree of 4 \cite{wu2019temporal}, combined with Triangel's metadata format. Overall, Prophet’s replacement policy, insertion policy, and Multi-path Victim Buffer contribute the most to performance speedup. Meanwhile, Prophet resizing proves most effective for workloads with relatively small metadata table requirements, such as sphinx3.

\textbf{Prophet replacement policy} is effective across most workloads by providing fine-grained management: it prioritizes the retention of metadata entries that achieve higher prefetching accuracy. Workloads with large working sets that are sensitive to cache pollution, such as mcf and omnetpp, particularly benefit from this feature, achieving performance gains of \mmi{14.53\%} and \mmi{9.89\%}, respectively. Additionally, the Prophet replacement policy also proves beneficial in reducing memory traffic for workloads like astar and omnetpp.

\textbf{Prophet insertion policy} is conservatively designed to avoid filtering out useful metadata (Section~\ref{subsec:step2}). Unlike the \textit{PatternConf} mechanism in Triangel, Prophet insertion policy avoids significant performance drops for omnetpp, soplex, and sphinx3. Meanwhile, it yields a significant performance speedup for mcf (\mmi{16.72\%}) and effectively reduces memory traffic across various workloads.

\textbf{Multi-path Victim Buffer} is developed to predict complex temporal patterns where a single memory address has multiple potential Markov targets, providing an opportunity to reduce more demand misses compared to the original metadata format design. This feature contributes to performance improvements across multiple workloads, with soplex showing a notable \mmi{13.46\%} performance speedup. \mmi{The Multi-path Victim Buffer only slightly increases memory bandwidth by 1.95\% due to our fine-grained management. Consequently, the Multi-path Victim Buffer enhances the performance of astar (which is sensitive to constrained memory bandwidth) rather than negatively impacting it.}

\textbf{Prophet Resizing} allocates metadata table space based on application requirements, allowing more space for the LLC without compromising metadata table capabilities. This approach yields performance benefits, especially for applications with smaller metadata needs. For example, sphinx3, which requires less than 1 MB of metadata table, achieves a \mmi{1.5\%} performance gain.

\textbf{The flexibility of Prophet.} Prophet's features are designed to be modular, allowing programmers to selectively enable or disable specific features based on evaluated performance and memory traffic. This adaptability means that if Prophet's impact on performance is unfavorable for certain workloads, programmers can selectively roll back to a subset of Prophet's features or revert to the runtime temporal prefetcher, such as gcc\_166.

\subsection{Storage Overhead}

The storage overhead of Prophet stems from its replacement state, the hint buffer, and the Multi-path Victim Buffer. 

\begin{itemize}
    \item \textbf{Prophet replacement stats.} Prophet supports a maximum metadata table of 1~MB, or 196,608 entries, with each entry requiring a 2-bit replacement state. Consequently, the total storage overhead for Prophet's replacement states is 48~KB.
    \item \textbf{Hint buffer.} If a hint buffer is used for processing hint information (Section~\ref{subsec:hint}), Prophet requires extra storage for this buffer. Our experiments indicate that a 128-entry hint buffer is sufficient for high performance, adding 0.19~KB.
    \item \textbf{Multi-path Victim Buffer} requires an additional 43 bits of storage per metadata entry: 31 bits for the memory address, 10 bits for the tag, and 2 bits for Prophet's replacement policy counter. For a buffer with 65,536 entries, this results in a storage overhead of 344~KB. We compare the performance gain of allocating this additional storage to the LLC, observing that the Multi-path Victim Buffer achieves an extra 2.21\% performance improvement (4.95\% vs. 2.74\%).
\end{itemize}


\subsection{\mmi{Energy Overhead}}
\label{suc:energy}

\mmi{We evaluate the energy overhead of Prophet with a focus on the memory hierarchy. We utilize CACTI~\cite{muralimanohar2009cacti} to model the energy consumption of the on-chip memory hierarchy under a $\qty{22}{nm}$ technology node, and estimate the DRAM access energy overhead to be 25$\times$ that of the LLC access overhead, similar to Triangel~\cite{ainsworth2024triangel}. Our experiments show that Prophet only introduces 1.6\% energy overhead for the memory hierarchy compared to Triangel. Given that Prophet's performance improvement over Triangel is 14.23\%, the 1.6\% energy overhead is relatively negligible.}
\section{\mmi{Related Works}}
\label{sec:related}

\mmi{In this section, we discuss the most relevant work in hardware temporal prefetching and profile-guided prefetching.}

\mmi{\textbf{\emph{Hardware temporal prefetchers.}} The most related work to Prophet is hardware temporal prefetchers \cite{jain2013linearizing, nesbit2004data, wu2019efficient, wenisch2009practical, wu2019temporal, bakhshalipour2018domino, ainsworth2024triangel, wu2021practical}. Like these prefetchers, Prophet requires metadata storage to maintain correlations between memory addresses. However, to the best of our knowledge, Prophet is the only approach that integrates profile-guided techniques into metadata storage management. By leveraging future knowledge, Prophet significantly optimizes metadata management while avoiding expensive hardware modifications. Furthermore, Prophet is fully compatible with existing hardware temporal prefetchers, allowing chip designers to flexibly choose between Prophet and hardware prefetchers (Section~\ref{subsec:breakdown}).}

\mmi{\textbf{\emph{Profile-guided prefetching.}} Profile-guided techniques have been applied to various aspects of prefetching, such as software prefetching \cite{ayers2020classifying, jamilan2022apt, zhang2024rpg2, khan2021dmon}, criticality-aware prefetching \cite{litz2022crisp}, and instruction prefetching \cite{khan2020spy}. Profile-guided software prefetching schemes identify prefetch kernels and utilize software prefetch instructions to predict them. Profile-guided criticality-aware prefetching focuses on identifying demand misses that lead to ROB stalls and extends the instruction scheduler to prioritize critical instructions. Profile-guided instruction prefetching schemes precisely identify I-cache misses and combine multiple non-contiguous prefetches into a single prefetch instruction.}

\mmi{Prophet sets it apart from these existing profile-guided prefetching schemes by focusing on a distinct aspect of prefetching: temporal prefetching, an area where prior techniques fall short. Additionally, Prophet introduces innovative methodologies for adaptable, lightweight, and compatible profile-guided optimizations, making it more seamless and practical to implement in industrial applications. }

\section{Conclusion}
\label{sec:conclusion}


In this paper, we propose Prophet, an adaptable, lightweight, and compatible profile-guided solution for temporal prefetching. Prophet injects hints into programs to guide the temporal prefetcher's replacement policy, insertion policy, and resizing operations, while dynamically tuning these hints to allow a single optimized binary to adapt to various program inputs. Our evaluations demonstrate that Prophet outperforms the state-of-the-art hardware temporal prefetcher and software indirect memory access prefetching scheme, with negligible profiling, analysis, and instruction overheads. Prophet demonstrates superior performance across a wide range of benchmarks and configurations.


\begin{acks}

We sincerely thank the anonymous reviewers for their insightful comments and suggestions. This research was supported by National Natural Science Foundation of China (NSFC) 62304192, Hong Kong Research Grants Council (RGC) YCRG Grant C6003-24Y, and ACCESS – AI Chip Center for Emerging Smart Systems, sponsored by the InnoHK initiative of the Innovation and Technology Commission of the Hong Kong Special Administrative Region Government. We thank HKUST Fok Ying Tung Research Institute and National Supercomputing Center in Guangzhou Nansha Sub-center for computational resources.

\end{acks}

\bibliographystyle{ACM-Reference-Format}
\bibliography{refs}


\end{document}